\documentclass{aa}  

\usepackage{graphicx}
\usepackage{txfonts}
\usepackage{subcaption}

\usepackage{lscape}

\usepackage{placeins}

\usepackage{gensymb}

\usepackage[colorlinks=true,linkcolor=blue,citecolor=blue]{hyperref}
\usepackage{multirow}

\begin{document}

   \title{Effect of galaxy mergers on dust masses}

   \author{W.~J.~Pearson\inst{\ref{inst:NCBJ}}
        \and M.~Romano\inst{\ref{inst:MPIR}, \ref{inst:INAF}}
        \and K.~Ma\l ek\inst{\ref{inst:NCBJ}}
        \and D.~Donevski\inst{\ref{inst:NCBJ}, \ref{inst:SISSA}}
        \and K.~Lisiecki\inst{\ref{inst:NCBJ}, \ref{inst:UVa}, \ref{inst:UVa2}}
        \and A.~P.~Cotter\inst{\ref{inst:NCBJ}}
        \and S.~Dey\inst{\ref{inst:NCBJ}}
        \and T.~Goto\inst{\ref{inst:NTHU}}
        \and S.~C.-C.~Ho\inst{\ref{inst:ANU}, \ref{inst:CAS}, \ref{inst:OzGrav}}
        \and D.~J.~Santos\inst{\ref{inst:DLSU}, \ref{inst:RTU}}
        }

   \institute{National Centre for Nuclear Research, Pasteura 7, 02-093 Warszawa, Poland\label{inst:NCBJ}\\\email{william.pearson@ncbj.gov.pl}
   \and Max-Planck-Institut f\"{o}r Radioastronomie, Auf dem H\"{u}gel 69, 53121, Bonn, Germany\label{inst:MPIR}
   \and INAF, OAPD, Vicolo dell'Osservatorio, 5, 35122 Padova, Italy\label{inst:INAF}
   \and SISSA, Via Bonomea 265, 34136 Trieste, Italy\label{inst:SISSA}
   \and Departamento de F\'{i}sica Te\'{o}rica, At\'{o}mica y \'{O}ptica, Universidad de Valladolid, 47011 Valladolid, Spain\label{inst:UVa}
   \and Laboratory for Disruptive Interdisciplinary Science (LaDIS), Universidad de Valladolid, 47011 Valladolid, Spain\label{inst:UVa2}
   \and Institute of Astronomy, National Tsing Hua University, No. 101, Section 2, Kuang-Fu Road, Hsinchu 30013, Taiwan\label{inst:NTHU}
   \and Research School of Astronomy and Astrophysics, The Australian National University, Canberra, ACT 2611, Australia\label{inst:ANU}
   \and Centre for Astrophysics and Supercomputing, Swinburne University of Technology, Hawthorn, VIC 3122, Australia\label{inst:CAS}
   \and OzGrav: The Australian Research Council Centre of Excellence for Gravitational Wave Discovery, Hawthorn, VIC 3122, Australia\label{inst:OzGrav}
   \and De La Salle University, 2401 Taft Avenue, Malate, Manila 1004, Philippines\label{inst:DLSU}
   \and Rizal Technological University, 704 Boni Avenue corner Sacrepante Street, Mandaluyong City, 1550 Metro Manila\label{inst:RTU}
   }

   \date{Received DD Month, 20XX}
 
  \abstract
   {}
   {Dust mass (M$_{Dust}$ ) is one of the key physical properties of galaxies that track the cold interstellar medium over cosmic history. However, the M$_{Dust}$ in merging galaxies is poorly understood. Here, we aim to derive how the cold dust mass of galaxies is affected by major mergers at low redshifts.}
   {M$_{Dust}$ was estimated for approximately 500 major mergers in the North Ecliptic Pole extragalactic field using four spectral energy distribution fitting runs with CIGALE, with different dust attenuation and emission models. We also analysed over 12\,000 mergers from the SIMBA simulation using its M$_{Dust}$. We matched merging and non-merging systems in stellar mass, redshift, and local density, to enable a controlled comparison of M$_{Dust}$. We then performed a second comparison using a control sample additionally matched in star-formation rate (SFR). Active galactic nuclei (AGN) host galaxies were then removed, and this matching repeated. Each CIGALE run, and SIMBA, were matched independently.}
   {Observed mergers have larger M$_{Dust}$ than the non-mergers by up to a factor of 1.82$\pm$0.35, or 1.60$\pm$0.44 without AGN-hosts. When matched by SFR, observed mergers show no significant change in M$_{Dust}$, both when including and excluding AGN-hosts. SIMBA galaxies show similar trends when AGN-hosts are included but there is still a slight increase in M$_{Dust}$ when matched by SFR. When AGN-hosts are excluded, the SIMBA mergers show smaller M$_{Dust}$ by a factor of  0.98$\pm$0.03, or 0.97$\pm$0.02 when matched in SFR. The higher the mass of a merger, the greater M$_{Dust}$ increase for observations for two of the CIGALE runs, or the smaller the decrease for SIMBA.}
   {Observed mergers are typically more dusty than their non-merging counterparts but only because they typically have higher SFRs, regardless of whether they host AGN or not. SIMBA mergers follow observations when AGN hosts are included, but show a reduction in M$_{Dust}$ when AGN hosts are excluded. Therefore, M$_{Dust}$ in SIMBA is linked, directly or indirectly, to AGN in a way that we do not see in observations.}

   \keywords{Galaxies: interactions --
            Galaxies: ISM --
            Galaxies: evolution --
            Galaxies: statistics
               }

   \maketitle
   \nolinenumbers

\section{Introduction}
Galaxy mergers form a key part of our current model of galaxy evolution and are a natural result of hierarchical growth found in the current cold dark matter cosmology \citep[e.g.][]{2014ARA&A..52..291C, 2015ARA&A..53...51S}. The most obvious impact of these interactions is the changes in morphology they cause. This can be the creation of tidal tails during the merging period, greater irregularity and asymmetry, the creation of galactic rings, or the creation of elliptical galaxies from spiral galaxies \citep[e.g.][]{2013ApJ...778...61T, 2024MNRAS.529..810R}.

During these morphological changes, material inside the merging galaxies is disturbed. Tidal forces are also known to push material into the super-massive black holes in the centre of galaxies, triggering active galactic nuclei \citep[e.g.][]{2020A&A...637A..94G, 2025arXiv250315317E, 2025OJAp....8E..12E}. The motion of gas is also understood to reduce the metallicity in the centres of merging galaxies as well as flatten the metallicity gradient within galaxies \citep[e.g.][]{2015MNRAS.451.4005G, 2018MNRAS.479.3381B, 2022MNRAS.509.2720S, 2024A&A...691A..19V, 2025ApJ...982..130P}. This motion can cause gas in the galaxies to become compressed and trigger star-formation, in some cases triggering such high star-formation rates (SFR) to be considered a starburst \citep[e.g.][]{2015MNRAS.454.1742K, 2019A&A...631A..51P, 2022A&A...661A..52P, 2024MNRAS.52711372A, 2025MNRAS.538L..31F}.

Observations of known merging systems at low redshift have found them to have almost twice the infrared (IR) luminosity than non-merging galaxies of similar stellar mass (M$_{\star}$) \citep{2007ApJ...660L..51L}. This increase in IR luminosity is believed to be a result of increased SFR during mergers \citep{2005AIPC..761..441G}, or due to tidal heating independent of SFR \citep{2015MNRAS.453..638D}. For the galaxy pairs (pre-mergers), the closer the two galaxies are, the greater the IR luminosity increase \citep{2007ApJ...660L..51L}. This is consistent with observations that find increasing SFR as the projected separation of galaxies decreases \citep[e.g.][]{2013MNRAS.433L..59P, 2022ApJ...940....4S, 2023MNRAS.522.5107B}. At higher redshifts, this picture is more complicated, with only 24\% of sub-millimetre galaxies at $z \approx 2.3$ being identified as merging \citep{2025ApJ...995...90C}.

More massive, dusty star-forming galaxies have a dust mass (M$_{Dust}$) that is correlated with SFR, meaning a galaxy with higher SFR will have a larger M$_{Dust}$ \citep[e.g.][]{2019ApJ...880..129P, 2020A&A...644A.144D}. However, the high SFRs found in starburst galaxies can result in the destruction or ejection of dust from a galaxy over short timescales \citep[20-25~Myr][]{2023MNRAS.521..662N}. The dust can then replenish itself over a few Myr once the SFR reduces back to typical levels.

The dust properties of galaxies have been found to be influenced by galaxy mergers. Dust is known to be heated in mergers, a likely result of increased SFR \citep{2005AIPC..761..441G}. Merging galaxies have also been shown to have greater dust attenuation than non-merging galaxies \citep[][]{2012A&A...548A.117Y}. This increase in dust attenuation becomes larger as the M$_{\star}$ of the galaxy increases. The relation between the IR excess (the ratio between the IR and far-ultraviolet emission, IRX) and the ultraviolet spectral slope \citep[$\beta$; e.g.][]{2021ApJ...913...37C, 2023A&A...679A..26H} is expected to break down for merging galaxies. The IRX-$\beta$ relation depends on the age of the stellar population, with older stars heating dust and increasing the IRX \citep{2018MNRAS.474.1718N, 2018ApJ...853...56R}. As mergers can induce star-formation, the age of the stellar population in the interacting galaxies can be reduced. Ultra-luminous infrared galaxies (L$_{IR} > 10^{12}$L$_{\odot}$; ULIRGS) are found to typically be merging galaxies and are IR bright due to dust heating understood to be a result of merger induced starbursts \citep[e.g.][]{1996ARA&A..34..749S}, or due to active galactic nuclei (AGN) activity increasing the dust temperature \citep[e.g.][]{2007ApJ...661L..13T}. These ULIRGS are also found to lie above the IRX-$\beta$ relation, thought to be due to their high M$_{Dust}$ \citep[e.g.][]{2010ApJ...715..572H, 2014ApJ...796...95C}. Simulations also find merging galaxies to lie above the IRX-$\beta$ relation \citep{2017ApJ...840...15S}.

AGN are also understood to influence the dust in their host galaxies. The winds driven by AGN activity can promote dust formation and growth in the centre of galaxies \citep{2002ApJ...567L.107E, 2019ApJ...885..126S}. Some of the dust formation channels around AGN are considered to be analogous to how dust is formed around Asymptotic Giant Branch stars \citep{2002ApJ...567L.107E, 2026arXiv260827571D}. The winds launched by AGN can also act to eject dust from galaxies, while the heating from the AGN accretion can also result in dust sublimation and destruction \citep[e.g.][]{2002ApJ...567L.107E, 2019ApJ...885..126S, 2025MNRAS.544.4532O}. Therefore, as mergers are believed to trigger AGN activity, these merger triggered AGN may have an impact on the M$_{Dust}$ found in merging galaxies.

Studies looking at M$_{Dust}$ in merging systems have found weak evidence of an increase in M$_{Dust}$ only for high mass (log(M$_{\star}$/M$_{\odot}$) $\geq 10.7$), spiral-spiral interactions and a decrease in M$_{Dust}$ for intermediate mass ($10.4 \leq$ log(M$_{\star}$/M$_{\odot}$) $<10.7$) spiral-elliptical mergers \citep{2016ApJS..222...16C}\footnote{\citet{2016ApJS..222...16C} present the change in gas mass (M$_{gas}$) but derive M$_{gas}$ from M$_{Dust}$ measurements with a fixed M$_{gas}$-to-M$_{Dust}$ ratio.}. Other mass ranges show no evidence of a change in M$_{Dust}$ in mergers compared to non-merging galaxies. For non-merging, quiescent galaxies, the high M$_{Dust}$ they can contain is understood to be partially a result of minor-mergers, mergers where the M$_{\star}$ ratio between the merging galaxies is less than four \citep[e.g.][]{2015MNRAS.449.3503D, 2016MNRAS.461.2856L, 2023A&A...678A..35D}.

In this work, we aim to study the link between galaxy major mergers and M$_{Dust}$, with both pre-mergers (merging pairs) and post-mergers treated collectively. We will determine whether M$_{Dust}$ follows the excess in SFR that we expect to find in merging galaxies. This is of particular interest as dust can be both enhanced through rapid enrichment and grain growth in dense gas, and depleted via shocks, outflows, or destruction \citep[e.g.][and referenced therein]{2025A&A...693A.118L}. Therefore, this work performs a controlled comparison of M$_{Dust}$ in merging and non-merging systems at fixed M$_{\star}$ and environment. We will also examine how AGN can influence any change seen by examining these changes when we include and exclude AGN-host galaxies.

This will be done observationally using known mergers in the North Ecliptic Pole (NEP) extragalactic field \citep{2022A&A...661A..52P}, due to its wealth of multi-wavelength data, as well as comparing the observational results to results from the SIMBA simulations \citep{2019MNRAS.486.2827D}, as these simulations internally model dust. In Section \ref{sect:data} we discuss the NEP and SIMBA data, Sect. \ref{sect:methods} discusses how we determine the change in M$_{Dust}$ in mergers, Sect. \ref{sect:results} presents our results, and Sect. \ref{sect:discussion} discusses these results. We conclude and summarise in Sect. \ref{sect:conclusion}.

\section{Data}\label{sect:data}
\subsection{North Ecliptic Pole}
The impact of major mergers on M$_{Dust}$ in galaxies was derived within the North Ecliptic Pole (NEP), covering an area of 5.4sq. deg. This field contains multi-wavelength data, from the UV to the sub-millimetre \citep{2021MNRAS.500.4078K}, that was taken as follow up observations to the AKARI NEP survey \citep{2012A&A...548A..29K}.

\subsubsection{Multi-wavelength catalogue}\label{sect:data:NEP}
In this work, we used $u$, $J$, and $Ks$ band observations from the Canada Fance Hawaii Telescope \citep[CFHT;][]{2014A&A...566A..60O}, $g$, $r$, $i$, $z$, and $Y$ band observations from the Hyper-Suprime Cam - NEP survey \citep[HSC-NEP;][]{2017PKAS...32..225G}, $J$, $H$ and $Ks$ band observations from Kitt Peak National Observatory \citep[KPNO;][]{2007AJ....133.2418I, 2014ApJS..214...20J}, $3.6\mu m$ and $4.5\mu m$ observations from \textit{Spitzer} InfraRed Array Camera \citep[IRAC;][]{2018ApJS..234...38N}, and $2.4 \mu m$, $3.2 \mu m$, $4.1 \mu m$, $7.0 \mu m$, $9.0 \mu m$, $11.0 \mu m$, $15.0 \mu m$, $18.0 \mu m$, and $24.0 \mu m$ band observations from AKARI \citep{2012A&A...548A..29K}, where these data are available.

We also used $100\mu m$ and $160\mu m$ band data from \textit{Herschel} Photodetector Array Camera and Spectrometer (PACS) and $250\mu m$, $350\mu m$, and $500\mu m$ band data from \textit{Herschel} Spectral and Photometric Imaging Receiver (SPIRE) \citep{2017PKAS...32..219P, 2019PASJ...71...13P, 2021MNRAS.507..129S}. These \textit{Herschel} data are known to suffer from confusion, and so we deblended these data with the XID+ tool \citep{2017MNRAS.464..885H} as described in Appendix \ref{app:deblend}. The far-infrared (FIR) data is needed to accurately constrain M$_{Dust}$ in galaxies as the peak of the thermal dust emission is in this wavelength region.

However, the redshifts of our galaxy sample is $z < 0.3$ (Sect. \ref{subsec:NEPmergers}). At these redshifts, if two pre-merger galaxies have projected separations of 20-30~kpc, this will be a spatial separation of 5-8~arcsec, within a Herschel SPIRE beam (17.6~arcsec at 250~$\mu$m). While our deblending has an average source density of approximately 2 sources per PSW beam (Appendix \ref{app:deblend}), it is possible that close pairs could be assigned incorrect flux densities.

To match these catalogues, we followed the method used by the Herschel Extragalactic Legacy Project \citep[HELP;][]{2021MNRAS.507..129S}. We correct the astrometry of the CFHT, HSC-NEP, KPNO, IRAC, and AKARI data by matching sources to GAIA positions, individually for each telescope, and subtracting the mean RA and Dec offsets. We then match the CFHT, KPNO, and IRAC sources to the HSC-NEP positions within 1~arcsec and match the AKARI data to sources within 1.5~arcsec of the HSC-NEP positions. The PACS and SPIRE sources were determined using the HSC-NEP positions, so do not require astrometry correction or positional matching.

SCUBA-2 850~$\mu$m is also available for 2sq. deg. of NEP \citep{2020MNRAS.498.5065S}, which can also be used to help constrain M$_{Dust}$. However, due to the large beam size ($\approx$ 15~arcsec) and the difficulty in deblending the SCUBA-2 data, we elect not to include this data in this work. The large beam size and poor ability to be deblended means we cannot reliably assign SCUBA-2 fluxes to our sources.

The photometric redshifts used were derived by \citet{2021MNRAS.506.6063H} using the CFHT, HSC-NEP, and IRAC data or by \citet{2021MNRAS.502..140H} using CFHT, HSC-NEP, KNPO, IRAC, AKARI, and Maidanak data. While \citet{2021MNRAS.502..140H} used more bands than \citet{2021MNRAS.506.6063H}, 26 compared to eight, we elected to use the \citet{2021MNRAS.506.6063H} redshifts for sources with photometric estimates from both catalogues. This is due to \citet{2021MNRAS.506.6063H} providing a significantly larger number of redshifts, and the redshifts estimated in \citet{2021MNRAS.506.6063H} were also used to select the merging galaxies. We refer the reader to \citet{2021MNRAS.506.6063H} and \citet{2021MNRAS.502..140H} for further details on the photometric redshifts. Where spectroscopic redshifts are available \citep{2013ApJS..207...37S, 2021MNRAS.502..140H, 2021MNRAS.500.4078K}, we used those in preference to photometric redshifts.

\subsubsection{NEP galaxy mergers}\label{subsec:NEPmergers}
The galaxy mergers used in this work come from \citet{2022A&A...661A..52P}. This catalogue was created through deep-learning and visual inspection and contains major mergers, that is mergers where the merging galaxies have M$_{\star}$ within a factor of four of each-other. The deep-learning neural network (NN) used the HSC-NEP $r$-band images and morphologies to identify the merger candidates, which were then visually inspected. Of the 34\,264 HSC-NEP galaxies at $z < 0.3$, 10\,195 were identified as merger candidates by the NN, of which 2109 were visually confirmed as mergers. These 2109 mergers were identified to be tidally disrupted and/or have a close companion in both projected separation and photometric redshift. Therefore, this sample contains both pre- and post-merger galaxies. We use the visually confirmed mergers for the rest of this work. Full details of the merger identification can be found in \citet{2022A&A...661A..52P}.

\subsubsection{NEP galaxy physical properties}
The physical properties of the galaxies, such as M$_{\star}$, SFR, and M$_{Dust}$, were estimated by fitting spectral energy distributions (SEDs) to the photometric data with Code Investigating GALaxy Emission \citep[CIGALE;][]{2019A&A...622A.103B}. We used an exponentially declining star-formation rate history (SFH) with an exponentially declining burst (\texttt{sfh2exp}), the \citet{2003MNRAS.344.1000B} stellar population (\texttt{bc03}) with a \citet{2003PASP..115..763C} initial mass function, continuum and line nebular emission (\texttt{nebular}), and the \citet{2016MNRAS.458.2288S} AGN emission model (\texttt{skitor2016}). The parameters used for these models can be found in Appendix \ref{app:models}.

As we are studying M$_{Dust}$ in galaxies, far-infrared (FIR) observations are needed to help constrain the emission from the dust. Therefore, we require that our sources have deblended detections in the PACS bands or deblended detections in the SPIRE bands. Where we only have deblended detections in the PACS bands, we also use an upper limit of 1.2~mJy in the 250$\mu$m SPIRE band (Appendix \ref{app:deblend}) during SED fitting. The requirement leaves 5941 galaxies, of which 531 are visually confirmed merging galaxies.

To ensure that our results are robust against using different dust models, we performed four different CIGALE runs using all combinations of the modified \citet[][\texttt{dustatt\_modified\_CF00}]{2000ApJ...539..718C} and modified \citet[][\texttt{dustatt\_modified\_starburst}]{2000ApJ...533..682C} dust attenuation models, and the  \citet[][\texttt{dl2014}]{2014ApJ...780..172D} and \citet[][\texttt{themis}]{2017A&A...602A..46J} dust emission models. Thus we matched each dust attenuation model with every dust emission model: \texttt{dustatt\_modified\_CF00} with \texttt{dl2014} (hereafter \texttt{cf00\_dr14}), \texttt{dustatt\_modified\_CF00} with \texttt{themis} (hereafter \texttt{cf00\_themis}), \texttt{dustatt\_modified\_starburst} with \texttt{themis} (hereafter \texttt{sb\_themis}), and \texttt{dustatt\_modified\_starburst} with \texttt{dl2014} (hereafter \texttt{sb\_dr14}. 

For the attenuation models, \texttt{dustatt\_modified\_CF00} uses two attenuation curves with different slopes: one for the interstellar medium (ISM) and one for birth clouds. Stars younger than 10~Myr are affected by both curves while older stars are only influenced by the ISM attenuation curve. The \texttt{dustatt\_modified\_starburst} attenuation, on the other hand, uses a single attenuation curve with a UV bump between the Lyman break and 150~nm \citep{2002ApJS..140..303L} for all stars. For the dust emission models, both are based on the \citet{2007ApJ...657..810D} emission model. The primary difference between the \texttt{dl2014} and \texttt{themis} models is that \texttt{dl2014} uses silicate-graphite-PAH (polycyclic aromatic hydrocarbon) based dust \citep{2014ApJ...780..172D, 2019A&A...624A..80N} while \texttt{themis} uses a silicate-core carbonaceous-mantle model \citep{2017A&A...602A..46J, 2019A&A...624A..80N}.

\begin{figure}
    \centering
    \resizebox{\hsize}{!}{\includegraphics{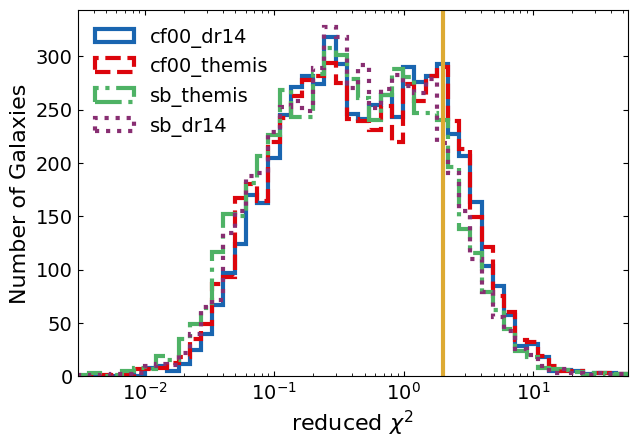}}
    \caption{Reduced $\chi^{2}$ distributions of the \texttt{cf00\_dr14} (solid blue line), \texttt{cf00\_themis} (dashed red line), \texttt{sb\_themis} (dot-dashed green line), and \texttt{sb\_dr14} (dotted purple line) CIGALE runs. The vertical yellow line is $\chi^{2} = 2$, the cut used in this work.}
    \label{fig:data:chi2}
\end{figure}

The parameters for the SFH, stellar population, and initial mass function (IMF) were kept the same for each CIGALE run and the parameters for the dust absorption and emission models were not changed between runs. The parameters used for the dust models can be found in Appendix \ref{app:models} along with example plots of the best fitting SEDs for four observed galaxies in Fig. \ref{fig:app:models:sed}. We only used results where the reduced $\chi^{2}$ of the CIGALE fitting is less than 2. As can be seen in Fig. \ref{fig:data:chi2}, the cut at $\chi^{2} = 2$ is reasonable for all four CIGALE runs. Our reduced $\chi^{2}$ cut leaves 82\%, 81\%, 86\%, and 86\% of the sources from the \texttt{cf00\_dr14}, \texttt{cf00\_themis}, \texttt{sb\_themis}, and \texttt{sb\_dr14} CIGALE runs, respectively.

The normalised local density ($\Sigma_{10}$) of the galaxies were taken from \citet{2021MNRAS.507.3070S}. These local densities are derived in the HSC-NEP catalogue using the 10$^{\mathrm{th}}$ nearest neighbour in projection, for galaxies within a redshift bin of width $0.65 \times (1+z)$ to account for photometric redshift uncertainties, and were normalised to the median density within the redshift bin. Edge corrections were applied to account for missing galaxies at the edges of the HSC-NEP survey. Full details of the local density estimates can be found in \citet{2021MNRAS.507.3070S}.

To calculate the 90\% M$_{\star}$ completeness limit for all four CIGALE runs, we follow the empirical method of \citet{2010A&A...523A..13P} applied to the r-band data \citep{2022A&A...661A..52P, 2024A&A...686A..94P}. This calculates the lowest observable M$_{\star}$ (M$_{lim}$) of a galaxy with an observed $r$-band magnitude ($r$) in a survey with a limiting $r$-band magnitude ($r_{lim}$) as:
\begin{equation}
    \mathrm{log}(\mathrm{M}_{lim}) = \mathrm{log}(\mathrm{M}_{\star}) - 0.4(r_{lim} - r).
\end{equation}
The 90\% completeness limit is then the value of M$_{lim}$ that 90\% of the faintest 20\% of sources have M$_{lim}$ below. Using this method, we find log(M$_{\star}$/M$_{\odot}$) limits of 7.90, 7.89, 7.59 and 7.60 for the \texttt{cf00\_dr14}, \texttt{cf00\_themis}, \texttt{sb\_themis}, and \texttt{sb\_dr14} runs, respectively. The SIMBA mass limit is log(M$_{\star}$/M$_{\odot}$)~$= 8.76$, below which SIMBA galaxies are no-longer resolved \citep{2019MNRAS.486.2827D}. As the SIMBA mass limit is the highest, we applied the mass limit of log(M$_{\star}$/M$_{\odot}$)~$> 8.76$ to each CIGALE run individually.

As AGN activity can contribute to M$_{Dust}$, we also create a second galaxy sample by removing likely AGN hosts by requiring the AGN contribution to the galaxy's luminosity to be less than 20\% (\texttt{fracAGN}$ < 0.2$). Again, this was done individually in each CIGALE run, removing 4422, 4540, 8322, and 8181 galaxies (either merging or non-merging) from the \texttt{cf00\_dr14}, \texttt{cf00\_themis}, \texttt{sb\_themis}, and \texttt{sb\_dr14} runs, respectively.

\subsubsection{CIGALE M$_{Dust}$ reliability}
To check that our M$_{Dust}$ estimates are reliable, we used CIGALE to generate a set of mock observations. These mock observations were generated by resampling the flux densities from the best fitting SEDs from a Gaussian distribution centred on the SED flux density and with a standard deviation of the uncertainty of the observed flux densities at that wavelength. CIGALE then fitted SEDs to these perturbed flux densities using the same models and model parameters.

\begin{figure*}
    \centering
    \resizebox{\hsize}{!}{\includegraphics{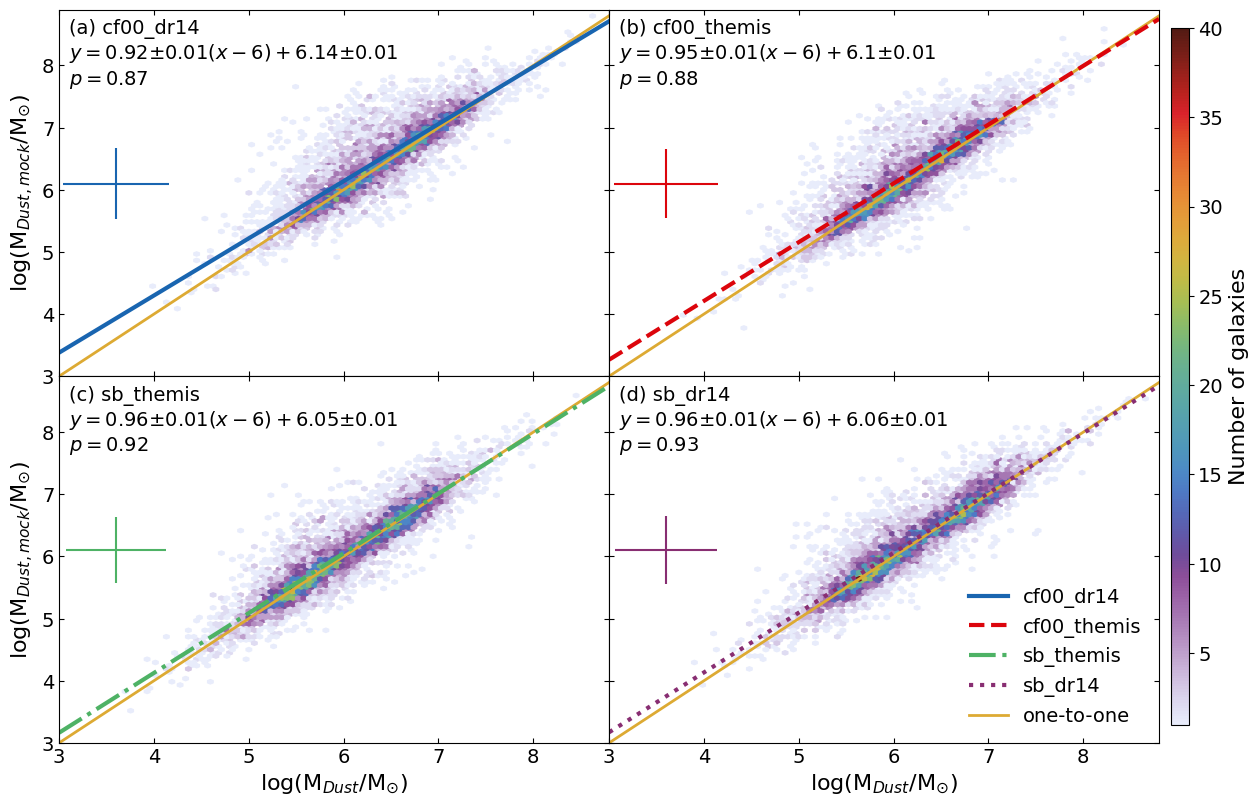}}
    \caption{Comparisons between the M$_{Dust}$ from observed flux densities and M$_{Dust}$ from mock observations (M$_{Dust,mock}$) for the \texttt{cf00\_dr14} (a, solid blue line), \texttt{cf00\_themis} (b, dashed red line), \texttt{sb\_themis} (c, dot-dashed green line), and \texttt{sb\_dr14} (d, dotted purple line) CIGALE runs. The colour corresponds to the number density from low (purple) to high (red). Overlaid are the lines when fitting the data to a line of the form $y = m(x-6) +c$, with the yellow line being the one-to-one line. The Pearson correlation coefficient ($p$) and the fitted values of $m$ and $c$ for each CIGALE run are shown in the top left of each panel. Typical uncertainties are shown as the coloured cross in each panel.}
    \label{fig:data:mdust}
\end{figure*}

We compare these mock M$_{Dust}$ values (M$_{Dust, mock}$) to the original estimates from CIGALE in Fig. \ref{fig:data:mdust}. Here we calculated the Pearson correlation coefficient ($p$) as well as fitting the data with a straight line of the form $y = m(x-6)+c$, with the location of normalisation chosen to be where we have data present (Fig. \ref{fig:data:mdust}). Perfectly reproduced M$_{Dust,mock}$ values will have $p = 1.0$, $m = 1.0$, and $c = 6.0$. As can be seen in Fig. \ref{fig:data:mdust}, all our CIGALE runs have $p$ greater than 0.87, $m$ values greater than 0.92 and $c$ values close to 6.0. Therefore, we are confident that our M$_{Dust}$ estimations are reliable.

\subsection{SIMBA}
We compared our observational results to M$_{Dust}$ in mergers and non-mergers from SIMBA's Flagship simulation \citep{2019MNRAS.486.2827D}. The flagship simulation we used includes both stellar and AGN feedback as well as on-the-fly dust lifecycle modelling. This SIMBA run simulated 1048$^{3}$ dark matter particles, with a mass of $9.6\times10^{7}$~M$_{\odot}$, and 1048$^{3}$ gas elements, with an initial mass of $1.82\times10^{7}$~M$_{\odot}$, in a box with width 100~cMpc~h$^{-1}$. The Flagship simulation used the Planck 2015 cosmology \citep{2016A&A...594A..13P} and a softening length of 0.5~ckpc~h$^{-1}$.

The SIMBA simulation also included a dust physics model \citep{2019MNRAS.490.1425L}. In SIMBA, dust grains are formed by supernova and asymptotic giant branch (AGB) stars, with dust growth in the ISM based on the models described in \citet{1998ApJ...501..643D} and \citet{2016MNRAS.457.3775M}. Destruction of dust is caused by supernova shocks and thermal plasma, following the models of \citet{2016MNRAS.457.3775M} and \citet{1995ApJ...448...84T}, respectively. This modelled dust formation from supernova and asymptotic giant branch (AGB) star ejecta following an updated version of the \citet{1998ApJ...501..643D} models. The simulations also allowed dust growth \citep{1995ApJ...448...84T} and destruction \citep{2016MNRAS.457.3775M}. This internal dust simulation is ideal for studying dust in the mergers found in SIMBA.

\subsubsection{SIMBA galaxy mergers}
To match the NEP merger sample, we used galaxies from snapshots 135 to 150 inclusive ($z < 0.3$), to match the redshift range of the observations. We selected galaxy mergers from the SIMBA simulation by selecting galaxies that either have merged in the last 300~Myr or will merge in the next 800~Myr \citep{2024A&A...687A..24M}. We also required these mergers to be major mergers, that is the merging galaxies or progenitor galaxies have masses within a factor of four of each other. These mass ratios were calculated in the snapshot where the secondary galaxy's M$_{\star}$ was most massive, following \citet{2015MNRAS.449...49R}. Non mergers were selected as galaxies that have not merged in the last 500~Myr and will not merge in the next 1000~Myr, following \citet{2024A&A...687A..45P}. This provided 15\,172 merging galaxies.

\subsubsection{SIMBA galaxy physical properties}
We calculated $\Sigma_{10}$ (the normalised local density) for galaxies in these snapshots by finding the distance to the 10$^{\mathrm{th}}$ nearest neighbour of each galaxy with the snapshot. If a galaxy's 10$^{\mathrm{th}}$ nearest neighbour is larger than the distance between the galaxy and the edge of the box, the 10$^{\mathrm{th}}$ nearest neighbour will likely be overestimated. Therefore, we removed these edge galaxies from further study. Once the edge galaxies were removed, we normalised the 10$^{\mathrm{th}}$nearest neighbour values by the median 10$^{\mathrm{th}}$ nearest neighbour in each snapshot to give $\Sigma_{10}$ for the SIMBA galaxies.

We also estimated the AGN contribution to the SIMBA galaxies' total luminosity. We calculated the AGN's bolometric luminosity (L$_{AGN}$) with the black hole accretion rate ($\mathrm{\dot{M}}_{BH}$), the black hole Eddington accretion limit ($\mathrm{\dot{M}}_{Edd}$), the Eddington fraction ($f_{edd} = \mathrm{\dot{M}}_{BH}/\mathrm{\dot{M}}_{Edd}$), and the radiative efficiency ($\epsilon_{r}$) following \citet{2022MNRAS.509.3015H}. For galaxies with $f_{edd} > 0.1$ we calculated L$_{AGN}$ as
\begin{equation}\label{eqn:high_fedd}
    \mathrm{L}_{AGN} = \frac{\epsilon_{r}}{1 - \epsilon_{r}} \mathrm{\dot{M}}_{BH}c^{2},
\end{equation}
where $c$ is the speed of light. For galaxies with $f_{edd} < 0.1$, we calculated L$_{AGN}$ as
\begin{equation}\label{eqn:low_fedd}
    \mathrm{L}_{AGN} = 10f_{Edd}\epsilon_{r}\mathrm{\dot{M}}_{BH}c^{2}.
\end{equation}
For both Eqn. \ref{eqn:high_fedd} and \ref{eqn:low_fedd}, we assume $\epsilon_{r} = 0.1$ \citep{2019MNRAS.486.2827D, 2022MNRAS.509.3015H}.

A galaxy's non-AGN luminosity (L$_{gal}$) was estimated by integrating the \citet{2003MNRAS.344.1000B} Padova 1994 simple stellar population (SSP) templates \citep{1991A&A...244...95A} with a \citet{2003PASP..115..763C} IMF. We matched the galaxies to the SSP template that most closely matched the galaxies' mass weighted stellar ages and metallicities provided by SIMBA, redshifted the wavelengths to match the redshift of the simulated galaxy, and integrated the emission to provide a luminosity. As the SSP templates provided by \citet{2003MNRAS.344.1000B} are normalised to 1~M$_{\odot}$, we multiply the luminosity from the integration by the M$_{\star}$ of the SIMBA galaxy, giving an approximate L$_{gal}$. The fraction of a galaxy's emission that is from the AGN ($F_{AGN}$) was then calculated as 
\begin{equation}
    F_{AGN} = \frac{L_{AGN}}{\mathrm{L}_{gal}+\mathrm{L}_{AGN}}.
\end{equation}
As with the observed galaxies, we create a second sample of galaxies where we removed the likely AGN hosts by requiring $F_{AGN} < 0.2$. The SSP templates cover the rest-frame wavelength range 91~\AA{} to 160~$\mu$m and so L$_{gal}$ is not the bolometric luminosity, meaning that $F_{AGN}$ are likely to be overestimated.

Other physical properties of the galaxies, such as M$_{\star}$, SFR, and M$_{Dust}$, were taken directly from the simulation.

\subsubsection{SIMBA FIR selected galaxies}

As our observed galaxies are selected through deblending the PACS or SPIRE data, we also need to select SIMBA galaxies that are likely to have similar FIR properties. To do this, we predicted the \textit{Herschel} PACS and SPIRE band fluxes using the CIGALE setup and parameters presented in Appendix \ref{app:deblend}. For this CIGALE run, we use the apparent magnitudes provided by SIMBA and try to closely match the observational data we have. We use CFHT $u$, Subaru Suprime-Cam \citep{2002PASJ...54..833M} $g$, $r$, $i$, and $z$, United Kingdom Infrared Telescope wide-field camera \citep{2007A&A...467..777C} $Y$, $J$, $H$, and $K$ bands, as well as \textit{Spitzer} IRAC 3.6$\mu$m and 4.5$\mu$m. Our observed galaxies have predicted PACS 100$\mu$m fluxes greater than 1~mJy or predicted SPIRE 250$\mu$m fluxes above 1.2~mJy estimated with the same CIGALE setup (Appendix \ref{app:deblend}). Therefore, we select our SIMBA galaxies to have predicted PACS 100$\mu$m fluxes greater than 1~mJy or predicted SPIRE 250$\mu$m fluxes above 1.2~mJy. This selection leaves 12\,875 merging galaxies.

\section{Methods}\label{sect:methods}

Due to our selection, all of the NEP galaxies were required to have log(M$_{\star}$/M$_{\odot}) > 8.76$, and the no AGN sample has the same criteria as well as \texttt{fracAGN}~$< 0.2$.  We matched each of our merging galaxies to five non-merging, control systems that have redshifts within 0.02 and $\Sigma_{10}$ within 0.02. We also required the non-merging control systems to have M$_{\star}$ within the M$_{\star}$ uncertainty of the merging galaxy; the mean log(M$_{\star}$/M$_{\odot}$) uncertainty across all four CIGALE runs is 0.22~dex. Matching in this way prevents any possibility that our results come from factors other than if the galaxies are merging or not. Following these criteria when including AGN-host galaxies left 107, 110, 127, and 124 merging galaxies from the \texttt{cf00\_dr14}, \texttt{cf00\_themis}, \texttt{sb\_themis}, and \texttt{sb\_dr14} runs, respectively. This matching for the no AGN sample left 68 mergers for \texttt{cf00\_dr14}, 72 mergers for \texttt{cf00\_themis}, 73 for \texttt{sb\_themis}, and 68 for \texttt{sb\_dr14} for further study.

To match the observations, all of the SIMBA galaxies were required to have log(M$_{\star}$/M$_{\odot}) > 8.76$, and again we create a no AGN sample by requiring $F_{AGN} < 0.2$. We matched each merging galaxy to five non-merging, control galaxies that are within the same snapshot and have $\Sigma_{10}$ within 0.02. We also required the control galaxies to have log(M$_{\star}$/M$_{\odot}$) within 0.22~dex, the mean M$_{\star}$ uncertainty of the NEP galaxies, of the merging galaxies' M$_{\star}$. This matching left 10\,590 merging galaxies that may host AGN and 8001 merging galaxies that do not host AGN, from SIMBA, for further study.

To further constrain our merger matching, we also matched each merging NEP galaxy to five control galaxies, as above, but also required the SFRs of the non-merging galaxies to be within the SFR uncertainty of the merging galaxy. For SIMBA, we required the log(SFR) to be within 0.55, the mean SFR uncertainty across all four CIGALE runs for the NEP galaxies. This stricter criteria left 65, 60, 52, and 52 \texttt{cf00\_dr14}, \texttt{cf00\_themis}, \texttt{sb\_themis}, and \texttt{sb\_dr14} mergers, respectively, from NEP and 9574 SIMBA mergers. When we exclude AGN-host galaxies, this drops to 44 \texttt{cf00\_dr14}, 44 \texttt{cf00\_themis}, 32 \texttt{sb\_themis}, and 31 \texttt{sb\_dr14} mergers from NEP as well as 7146 SIMBA mergers.

To study the change in M$_{Dust}$, we defined this change ($\Delta$Dust) for each merging galaxy to be:
\begin{equation}
    \Delta \mathrm{Dust}_{i} = \log\bigg(\frac{\mathrm{M}_{Dust,mgr,i}}{M_{\odot}}\bigg) - median_{j=1,...,5}\bigg[ \log\bigg(\frac{\mathrm{M}_{Dust,nmgr,j}}{M_{\odot}}\bigg) \bigg],
\end{equation}
where M$_{Dust,mgr,i}$ is the dust mass of the $i^{\mathrm{th}}$ merging galaxy and M$_{Dust,nmgr,j}$ is the dust mass of the $j^{\mathrm{th}}$ non-merging control galaxy that was matched to the $i^{\mathrm{th}}$ merging galaxy.

\begin{figure}
    \resizebox{\hsize}{!}{\includegraphics{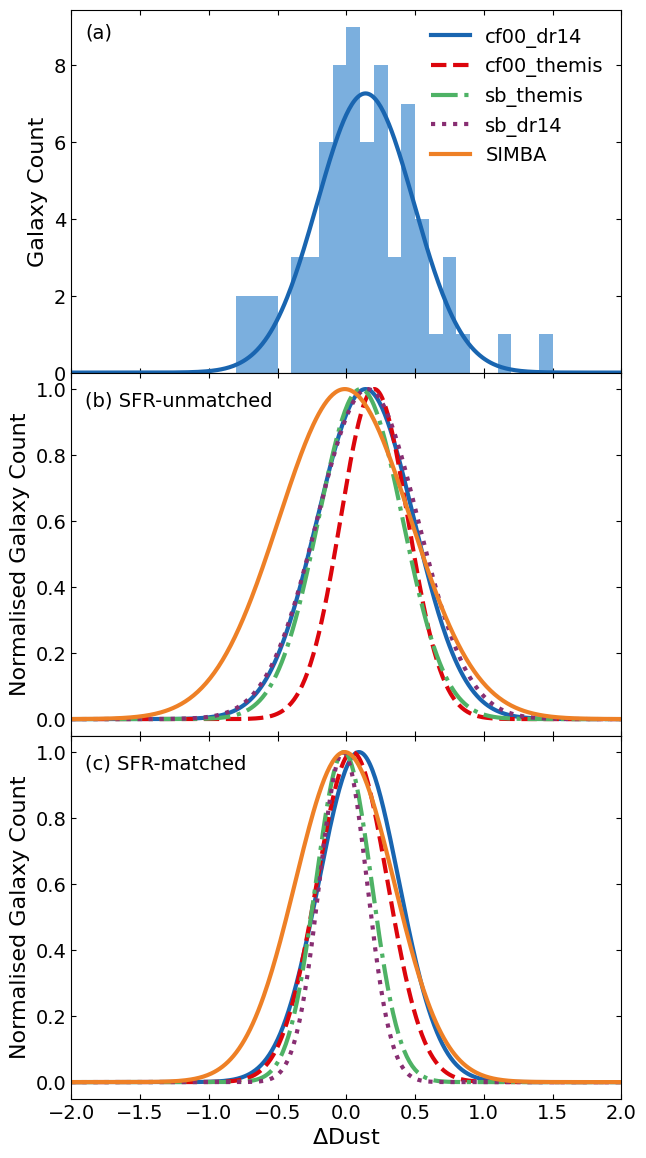}}
	\caption{(a) Example of fitting a Gaussian distribution to the distribution of $\Delta$Dust for the \texttt{cf00\_dr14} CIGALE run with control galaxies that are not matched in SFR to the merging galaxies and excluding AGN hosts from mergers and non-mergers. (b) All Gaussian fitting results for \texttt{cf00\_dr14} (solid blue line), \texttt{cf00\_themis} (dashed red line), \texttt{sb\_themis} (dot-dashed green line), and \texttt{sb\_dr14} (dotted purple line) CIGALE runs and SIMBA (solid orange line) with control galaxies that are not matched in SFR and excluding AGN hosts from mergers and non-mergers (c) As panel b but with control galaxies that are matched in SFR. As SIMBA contains two orders of magnitude more merging galaxies, panels b and c are normalised to the maximum of each distribution.}
	\label{fig:method:distribution}
\end{figure}

With the $\Delta$Dust for the merging galaxies, we then fitted a Gaussian distribution to the distribution of $\Delta$Dust, as shown in Fig. \ref{fig:method:distribution}. We performed this process for all four CIGALE runs and the SIMBA galaxies, both when the mergers were and were not matched by SFR. We then used the mean of the Gaussian distribution ($\mu$) as the change in dust mass for the population ($\mu\Delta$Dust). The uncertainties in $\mu$ were derived by bootstrapping, where we recalculated $\Delta$Dust after resampling the dust masses of the merging and non-merging galaxies from a normal distribution centred on the dust mass from CIGALE (or SIMBA) and with a standard deviation of the uncertainty from CIGALE (or the mean uncertainty across all four CIGALE runs, 0.55~dex, for SIMBA galaxies). We fitted a Gaussian distribution to these resampled $\Delta$Dust values. The standard deviation of the $\mu$ values from 1000 such resamplings was taken as the uncertainty on the measured $\mu$.

\section{Results}\label{sect:results}
\begin{figure}
    \resizebox{\hsize}{!}{\includegraphics{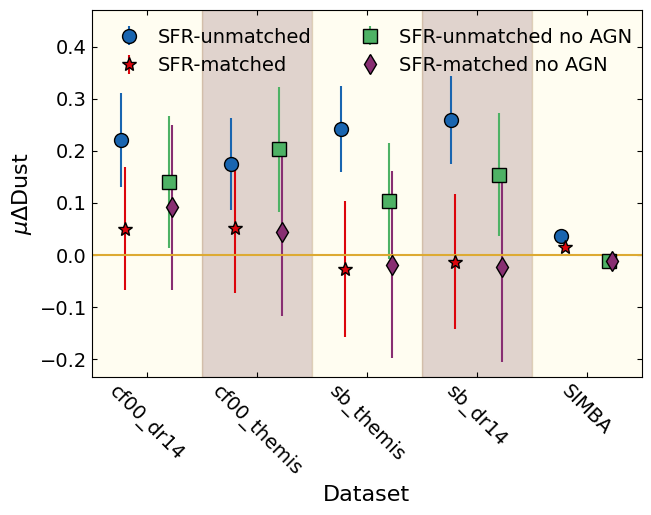}}
	\caption{$\mu\Delta$Dust for the four CIGALE runs (\texttt{cf00\_dr14}, \texttt{cf00\_themis}, \texttt{sb\_themis}, \texttt{sb\_dr14}) and the SIMBA galaxies. Blue circles show $\mu\Delta$Dust when the non-merger control galaxies are matched to the mergers without considering SFRs. Red stars show $\mu\Delta$Dust when the non-merger control galaxies are selected with consideration to the merging galaxies' SFRs. Green squares show $\mu\Delta$Dust when excluding AGN-host galaxies before matching and the non-merger control galaxies are matched to the mergers without considering SFRs. Purple diamonds show $\mu\Delta$Dust when excluding AGN-host galaxies before matching and the non-merger control galaxies are selected with consideration to the merging galaxies' SFRs. The yellow horizontal line indicates where $\mu\Delta$Dust~$= 0$. Uncertainties are from bootstrapping $\mu\Delta$Dust.}
	\label{fig:results:main}
\end{figure}

We find that $\Delta$Dust is typically higher in merging galaxies than non-merging galaxies for all four CIGALE runs, both when including and excluding AGN hosts, as shown in Fig. \ref{fig:results:main}. However, the exact enhancement is dependent on the absorption and emission models used as well as if the AGN are included or not. The exception is for the \texttt{sb\_themis} run when we exclude AGN hosts, which is just within error of $\mu\Delta$Dust~=~0. We also find that SIMBA merging galaxies are more dusty than non-merging galaxies when we include AGN hosts. On the other hand, we find that $\mu\Delta$Dust is lower for merging galaxies than non-merging galaxies in SIMBA when we exclude AGN-host galaxies, although this is within error of there being no difference between mergers and non-mergers.

For the merging galaxies that are also matched to non-merging galaxies with similar SFR, we find that all CIGALE runs are consistent with $\mu\Delta$Dust~$= 0$, although both runs using \texttt{dustatt\_modified\_CF00} have a weak indication that the M$_{Dust}$ in mergers is larger than non-mergers. The two CIGALE runs with \texttt{dustatt\_modified\_starburst}, on the other hand, have a weak indication that merging galaxies are less dusty. This is for both when we do and when we do not allow AGN-host galaxies. The SIMBA galaxies again show a positive $\mu\Delta$Dust when we include AGN host galaxies and a negative $\mu\Delta$Dust when AGN hosts are excluded. This shows that SIMBA merging galaxies without AGN are less dusty than their non-merging counterparts when matched in SFR. We summarise our results in Fig. \ref{fig:results:main} and Table \ref{table:results:main}.

\begin{table}
	\caption{$\mu\Delta$Dust values for the four CIGALE runs and SIMBA galaxies when unmatched and matched in SFR and when including and not including AGN-host galaxies.}
	\label{table:results:main}
	\centering
	\begin{tabular}{cccc}
		\hline
        \hline
        \multirow{2}{*}{Dataset} & \multicolumn{2}{c}{$\mu\Delta$Dust} & \multirow{2}{*}{AGN}\\
        & Unmatched SFR & Matched SFR & \\
        \hline
        \texttt{cf00\_dr14} & 0.221$\pm$0.090 & 0.051$\pm$0.118 & yes \\
        \texttt{cf00\_themis} & 0.175$\pm$0.088 & 0.052$\pm$0.126 & yes \\
        \texttt{sb\_themis} & 0.242$\pm$0.083 & -0.027$\pm$0.131 & yes \\
        \texttt{sb\_dr14} & 0.259$\pm$0.084 & -0.013$\pm$0.130 & yes \\
        SIMBA & 0.036$\pm$0.009 & 0.016$\pm$0.009 & yes \\
        \hline
        \texttt{cf00\_dr14} & 0.140$\pm$0.126 & 0.092$\pm$0.159 & no \\
        \texttt{cf00\_themis} & 0.203$\pm$0.120 & 0.044$\pm$0.161 & no \\
        \texttt{sb\_themis} & 0.104$\pm$0.111 & -0.018$\pm$0.179 & no \\
        \texttt{sb\_dr14} & 0.155$\pm$0.119 & -0.023$\pm$0.184 & no \\
        SIMBA & -0.011$\pm$0.012 & -0.012$\pm$0.010 & no \\
        \hline
	\end{tabular}
\end{table}

\begin{figure*}
    \resizebox{\hsize}{!}{\includegraphics{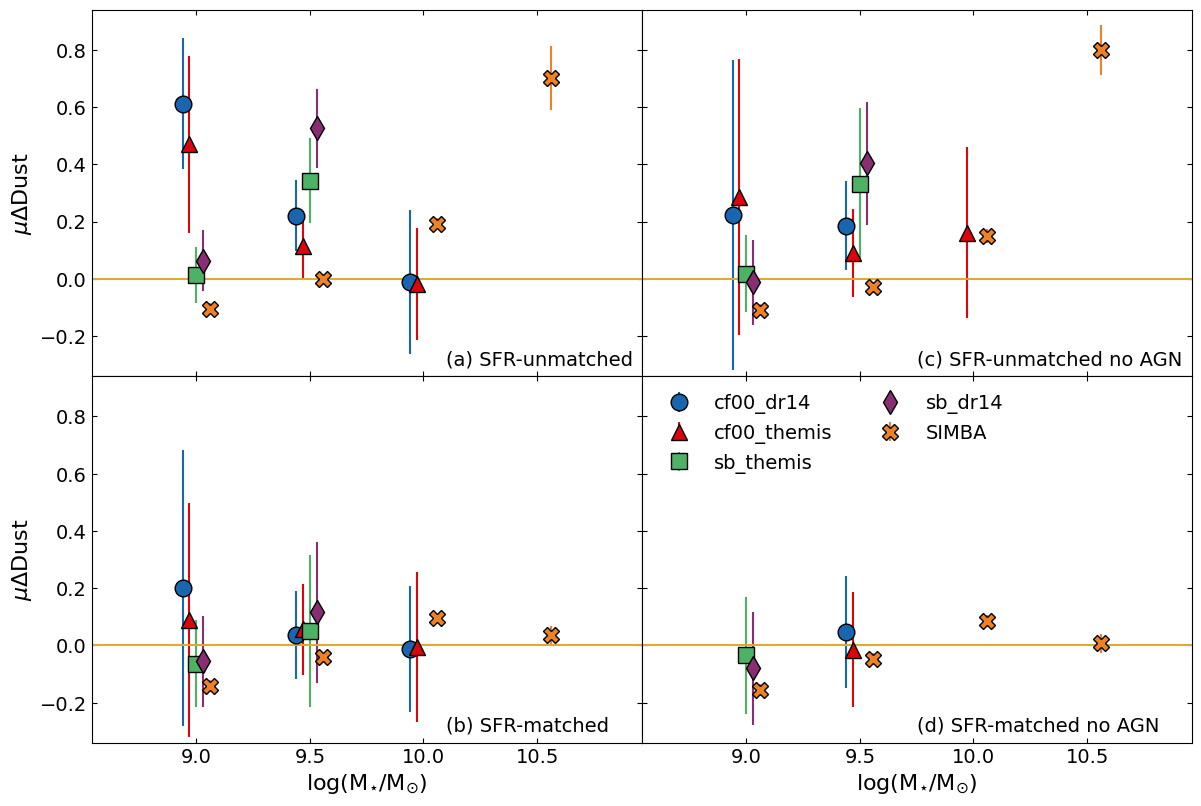}}
	\caption{$\mu\Delta$Dust for the four CIGALE runs (\texttt{cf00\_themis} as red triangles, \texttt{sb\_themis} as green squares, \texttt{sb\_dr14} as purple diamonds) and the SIMBA galaxies (orange crosses) when binned by M$_{\star}$. Panel (a) shows the results when the non-merger control galaxies are matched to the mergers without considering SFRs. Panel (b) shows $\mu\Delta$Dust when the non-merger control galaxies are selected with consideration to the merging galaxies' SFRs. Panels (c) and (d) are as panels (a) and (b) but when we exclude AGN-host galaxies before matching. The yellow horizontal line indicates where $\mu\Delta$Dust~$= 0$. Uncertainties are from bootstrapping $\mu\Delta$Dust.}
	\label{fig:results:bymass}
\end{figure*}

We repeated the analysis after binning the merging galaxies by M$_{star}$, into bins of width log(M$_{\star}$/M$_{\odot}$) = 0.5. Again, we fitted Gaussian distributions to the $\Delta$Dust values within each M$_{\star}$ bin, this time requiring that each bin contained at least 10 merging galaxies. We present these results in Fig. \ref{fig:results:bymass}, including when we do not match in SFR (upper row) and do match in SFR (lower row) as well as when we include AGN-host galaxies (left column) and do not include AGN-host galaxies (right column).

For the SFR-unmatched results in \ref{fig:results:bymass}a and c, we find that the two CIGALE runs using \texttt{dustatt\_modified\_starburst} have $\mu\Delta$Dust that increases as M$_{\star}$ increases. However due to the low number of merging galaxies with $\mathrm{log(M}_{\star}/\mathrm{M}_{\odot}) > 9.75$, this is only seen for lower masses. The other two CIGALE runs, using \texttt{dustatt\_modified\_CF00}, see a decreasing $\mu\Delta$Dust with increasing M$_{\star}$ when including AGN-host galaxies (Fig. \ref{fig:results:bymass}a) but no significant trend when the AGN-hosts are excluded (Fig. \ref{fig:results:bymass}c). The lowest mass galaxies, $8.75 < \mathrm{log(M}_{\star}/\mathrm{M}_{\odot}) \leq 9.25$, have $\mu\Delta$Dust values consistent with no M$_{Dust}$ increase in the two \texttt{dustatt\_modified\_starburst} runs when including and excluding AGN-hosts. This is also seen for the two \texttt{dustatt\_modified\_CF00} runs when we exclude AGN-host galaxies.

When we matched the SFR of the control galaxies to the SFR of the observed merging galaxies, there is still a very weak indication that there is an increase in M$_{Dust}$ in merging galaxies as M$_{\star}$ increases in Fig. \ref{fig:results:bymass}b for the two CIGALE runs using \texttt{dustatt\_modified\_starburst}. However, as the uncertainties are larger than the change in $\mu\Delta$Dust, this assertion is very optimistic. There is also a very weak indication that there is a decrease in $\mu\Delta$Dust as M$_{\star}$ increases for the two runs using \texttt{dustatt\_modified\_CF00} but, again, the large uncertainties prevent any meaningful statement from being made. Due to the low number of merging galaxies that do not host AGN and are SFR-matched to the controls, each of the four CIGALE runs only has one M$_{\star}$ bin that $\mu\Delta$Dust can be calculated for and so no trend can be seen for any CIGALE run in Fig. \ref{fig:results:bymass}d. All of these values of $\mu\Delta$Dust that we present are consistent with no change in M$_{Dust}$.

For the SIMBA galaxies, $\mu\Delta$Dust is less than, or consistent with, 0 at $\mathrm{log(M}_{\star}/\mathrm{M}_{\odot}) \leq 9.75$ and continues to rise out to $\mathrm{log(M}_{\star}/\mathrm{M}_{\odot}) = 10.25$ regardless of SFR matching or including AGN-host galaxies. This rise is continued out to the highest mass bin when we are not matching the control galaxies to the merging galaxies' SFR. When we do match in SFR, $\mu\Delta$Dust drops when we move into the highest mass bin.

Our analysis was repeated on a sample that is nearly ten times larger but primarily contains galaxies without FIR detections. The results we find are qualitatively consistent with what we find when only using \textit{Herschel} detected sources, have uncertainties that are much smaller, but M$_{Dust}$ that are likely to be less reliable. These results for when we also include non-\textit{Herschel} detected sources are presented in Appendix \ref{app:notFIRonly}.

\section{Discussion}\label{sect:discussion}
\subsection{Dust enhancement in merging galaxies}
\subsubsection{Control galaxies unmatched in SFR}
When we do not match the SFR of the control galaxies to the SFR of the observed merging galaxies, we find that the merging galaxies typically have higher M$_{Dust}$ than the non-merging galaxies. This would suggest there is some mechanism that either produced dust during the merger or, alternatively, prevents dust destruction. In massive, dusty, star-forming galaxies, the dust mass is seen to increase with SFR \citep[e.g.][]{2019ApJ...880..129P, 2020A&A...644A.144D}. With the SFR typically found to increase in merging galaxies \citep[e.g.][]{2015MNRAS.454.1742K, 2019A&A...631A..51P, 2022A&A...661A..52P, 2024MNRAS.52711372A, 2025MNRAS.538L..31F}, we would expect to see merging galaxies to be more dusty than non-merging galaxies. This assumes that the galaxies were dusty before they started to merge.

Tidal interactions in mergers are known to funnel material into the centre of galaxies, increasing the density of gas in the galactic centres. Higher gas densities have been shown to have higher dust growth rates through accretion \citep{2020MNRAS.499.6035M}. Therefore, the higher central gas densities caused by a merger may also be causing an increase in M$_{Dust}$ in our merging galaxies. This assumes that the galaxies were not gas poor before they started to merge.

With the $\mu\Delta$Dust increasing with M$_{\star}$ (Fig. \ref{fig:results:bymass}), this would align with the expectation that more massive, star-forming galaxies have larger M$_{Dust}$. At higher M$_{\star}$, we would expect the control galaxies that are not matched in SFR to have lower SFRs than their merging counterparts as quiescent galaxies are more prevalent \citep[e.g.][]{2010ApJ...721..193P,2023A&A...677A.184W}. This increase in $\mu\Delta$Dust is seen in both observations using \texttt{dustatt\_modified\_starburst} and SIMBA.

On the other hand, high SFRs in starburst galaxies can result in the destruction or ejection of dust over short timescales, 20-25~Myr \citep{2023MNRAS.521..662N}. Therefore for the highest SFR merging galaxies we may expect to see a reduction in M$_{Dust}$ at fixed M$_{\star}$. However, as merging galaxies are not typically found to be star-bursting \citep[e.g.][]{2015MNRAS.454.1742K, 2019A&A...631A..51P, 2022A&A...661A..52P, 2022ApJ...940....4S}, and the merger sample used in this work is not primarily star-bursting \citep{2022A&A...661A..52P}, we would only expect the minority of galaxies to experience this dust destruction or ejection.

AGN are also known to be an avenue for dust ejection through galactic outflows \citep[e.g.][]{2019ApJ...885..126S}. However, there is a weak indication that when we remove AGN-host galaxies from our sample, $\mu\Delta$Dust reduces: the values of $\mu\Delta$Dust are reduced when we remove AGN hosts but the change is typically within error. This hints that AGN are an avenue for dust creation in galaxy mergers and are not primarily a source of dust removal.

When we have AGN-host galaxies from SIMBA, the simulation results are qualitatively similar to the observations: there is a clear increase in M$_{Dust}$ in merging galaxies in Fig. \ref{fig:results:main}. However, unlike the observations, when we remove the AGN-host galaxies, merging galaxies are found to have less dust than their non-merging companions. This shows that in SIMBA, the AGN are intertwined with the physical properties of the simulated mergers in a way that differs from the observations.

For the SIMBA merging galaxies to have lower M$_{Dust}$ than the non-merging counterparts once AGN hosts are removed, it may be that this dust destruction in the simulation is triggered earlier than in observations. The peak of the log of the specific SFR (SFR/M$_{\star}$, sSFR) is between -9.50 and -9.25 for the SIMBA galaxies compared to between -9.75 and -9.50 for all four CIGALE runs. This difference is small, within the typical SFR uncertainty of 0.55~dex for the CIGALE SFR estimates. Therefore, there is not a systematically high SFR in the SIMBA galaxies that may be the cause of the dust destruction.

After removing AGN-host galaxies, SIMBA galaxies are still affected by feedback through radiative heating and outflows. These feedback mechanisms are known to be powerful enough to prevent efficient dust growth due to ISM heating. Previous studies have found that SIMBA galaxies with log(M$_{\star}$/M$_{\odot}$)$ > 9.75$ contain a large number of young and metal-rich stars that can balance dust growth in the ISM and dust removal due to supernovae shocks \citep{2025A&A...693A.118L}. Similar process have been seen in observations, which attribute these changes to age-dependent evolution \citep{2017MNRAS.465...54C, 2025A&A...694A..82S}. Therefore, finding positive $\mu\Delta$Dust values for SIMBA galaxies with log(M$_{\star}$/M$_{\odot}$)$ > 9.75$ is unsurprising.

To add to this, dust production at lower redshifts in SIMBA is dominated by dust growth in the ISM. If galaxies have lower gas fraction, a results of enhanced SFR during a merger, they will need 200-400~Myr for the dust to reform. The dust destruction by supernovae is on a much shorter timescale, 50-100~Myr \citep{2020A&A...644A.144D}. Our merger selection criteria selects post-merger galaxies that have merged in the last 300~Myr. If we expect a burst of star-formation at coalescence \citep{2025MNRAS.538L..31F, 2025A&A...704A.234P}, this 300~Myr timescale does not give enough time for the dust to be reformed before we no-longer consider the galaxies to be merging. For the pre-mergers in our SIMBA sample, the most recent close passage is likely to be within the last few hundred Myr \citep{2024MNRAS.529.1493P}. With an increase in SFR likely to occur during close passages \citep[e.g.][]{2019MNRAS.485.1320M, 2025MNRAS.538L..31F}, these galaxies will again not have had enough time for the dust to reassert itself. This longer timescale for dust production compared to destruction may be why we do not see an increase in $\mu\Delta$Dust in SIMBA galaxies when we remove AGN hosts.

\subsubsection{Matched in SFR}
To remove any potential SFR dependence on M$_{Dust}$, we also matched our merging galaxies to control galaxies with similar SFR (Fig. \ref{fig:results:main}, red stars). For the sample of galaxies including AGN hosts, all the CIGALE runs have $\mu\Delta$Dust values that drop to become consistent with zero. For the two runs using \texttt{dustatt\_modified\_CF00} there is a weak indication that $\mu\Delta$Dust is positive, and hence a weak indication that mergers are more dusty than non-mergers, even if we correct for SFR. The other two runs have a weak indication that $\mu\Delta$Dust is negative, suggesting the opposite. When we exclude AGN hosts, again all four CIGALE runs have $\mu\Delta$Dust that decreases to become consistent with zero. As with the sample including AGN hosts, the \texttt{dustatt\_modified\_CF00} runs show a weak indication that $\mu\Delta$Dust is positive while the the \texttt{dustatt\_modified\_starburst} runs show a weakly negative $\mu\Delta$Dust. We note that the $\mu\Delta$Dust when we match the SFR and include AGN hosts for the  \texttt{cf00\_themis} run is consistent, within uncertainty, with the result of the same run when we do not match by SFR. The $\mu\Delta$Dust for the SFR matched and unmatched \texttt{cf00\_dr14}, \texttt{sb\_themis}, and \texttt{sb\_dr14} runs are also consistent, within error, when we exclude AGN hosts.

Therefore, for the observed galaxies, the increase in M$_{Dust}$ seen before is likely due to the higher typical SFR of a merging galaxy compared to a non-merging system. The lack of change when we include or exclude AGN suggests that the dust in merging galaxies that do not host AGN is not fundamentally different to dust in non-AGN host non-mergers. A larger sample of merging galaxies, and hence smaller uncertainties, will make it more apparent if the enhancement or decrement in M$_{Dust}$ for SFR-matched galaxy mergers is truly present.

The SIMBA galaxies also show a decrease in $\mu\Delta$Dust when we required the control galaxies to have similar SFR to the merging galaxies. When we include AGN-host galaxies, the $\mu\Delta$Dust value drops but remains above zero. This partially aligns with the results from the \texttt{dustatt\_modified\_CF00} CIGALE runs: we see a drop and the value of $\mu\Delta$Dust remains positive. However, the SIMBA galaxies do not have $\mu\Delta$Dust consistent with zero, likely a result of a much larger number of galaxies reducing the uncertainty on $\mu\Delta$Dust.

When we remove AGN-host galaxies, if SIMBA does trigger dust destruction at lower SFRs than we find in observations, we would expect $\mu\Delta$Dust to increase when the control galaxies are matched by SFR compared to the SFR-unmatched sample. This not being seen suggests that there is another physical process within the merging galaxies that is acting to either remove the dust, prevent its creation, or both. SIMBA does not only explicitly model dust destruction but also models the ISM heating \citep{2019MNRAS.486.2827D}. This heating can indirectly slow down the dust growth timescale, hindering the creation of new dust despite higher SFRs. This does not prevent the negative $\mu\Delta$Dust for the SFR-unmatched sample being a result of dust destruction in SIMBA, but indicates that there may be something else unphysical happening in the simulation. However, a negative $\mu\Delta$Dust once matched in SFR is consistent with what we see with the \texttt{dustatt\_modified\_starburst} CIGALE runs. Again, the much smaller uncertainty seen with the SIMBA galaxies is likely a result of a larger number of merging galaxies.

Despite our removal of AGN-host galaxies from SIMBA, high-mass galaxies in SIMBA will still experience torque-limited feedback mechanisms that prevent dust growth \citep{2025A&A...693A.118L}. These feedback mechanisms are normally associated with the presence of AGN. This is a likely explanation for why we observe a drop in $\mu\Delta$Dust for the highest mass SIMBA galaxies in Fig. \ref{fig:results:bymass}d, when the control galaxies are matched in SFR to the merging galaxies. It is also a likely an explanation for why we see a similar trend when we include AGN-host galaxies in Fig. \ref{fig:results:bymass}b.

\subsection{CIGALE dust models}
As the assumed dust emission models, and to a lesser extent the dust attenuation modes, will influence M$_{Dust}$, we used two dust models and two attenuation models, providing us the four CIGALE runs. For the SFR-unmatched controls, all of the four runs are consistent within error (Fig. \ref{fig:results:main} and Table \ref{table:results:main}). This is seen both when we include and exclude AGN-host galaxies. Neither the dust attenuation or dust emission models appear to have a notable influence on $\mu\Delta$Dust.

When we match the SFR of the galaxies, we see a slightly different picture. All four CIGALE runs again have $\mu\Delta$Dust values that are consistent with one another. Again, this is seen when we include and exclude AGN hosts. There is a weak indication that the \texttt{dustatt\_modified\_starburst} attenuation model produces slightly lower M$_{Dust}$ in merging galaxies than \texttt{dustatt\_modified\_CF00}. As this difference is within error, we would require more data to better constrain the uncertainty and to make this assertion more than conjecture.

With all four CIGALE runs being in agreement with one-another, there is no strong evidence that the dust emission or attenuation models have an influence on our results. However, with the hints of disagreements seen between the different dust attenuation models, a larger sample of merging galaxies would be needed to reduce the uncertainties of $\mu\Delta$Dust. This would then allow us to be certain that there are no differences in our results due to the models used.

\subsection{Control SFR matching}
\begin{figure}
    \resizebox{\hsize}{!}{\includegraphics{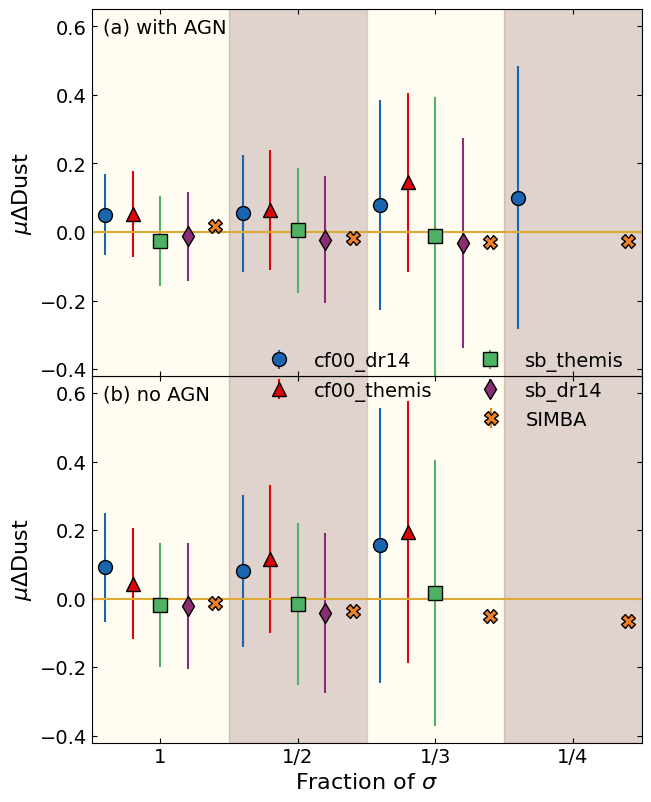}}
	\caption{$\mu\Delta$Dust for the four CIGALE runs (\texttt{cf00\_dr14} as blue circles, \texttt{cf00\_themis} as red triangles, \texttt{sb\_themis} as green squares, \texttt{sb\_dr14} as purple diamonds) and the SIMBA galaxies (orange crosses) as the fraction of the merging galaxy SFR uncertainty ($\sigma$) used to match the control galaxies becomes smaller. Panel (a) is for the full galaxy samples and panel (b) is where we have removed AGN hosts before matching to mergers to non-mergers. The yellow horizontal line indicates where $\mu\Delta$Dust~$= 0$. Uncertainties are from bootstrapping $\mu\Delta$Dust.}
	\label{fig:discussion:sfr_matching}
\end{figure}

As seen in Fig. \ref{fig:method:distribution}, the width of the Gaussian distributions is smaller when the control galaxies are matched in SFR than when they are not, meaning that there is less spread in $\mu\Delta$Dust when the SFRs are matched. Figure \ref{fig:results:main} also typically shows a decrease in $\mu\Delta$Dust when the SFRs of the control galaxies are matched to the SFR of the merging galaxies. To see if more closely matching the SFR of the control galaxies to the SFR of the merging galaxies further reduces $\mu\Delta$Dust, we re-selected control galaxies. This was done by requiring the SFR of the control galaxies to be within 1/2, 1/3, or 1/4 times the SFR uncertainty of the merging galaxy ($\sigma$). We then fitted the Gaussian distributions, as done in Sect. \ref{sect:methods}, requiring at least 10 merging galaxies to be matched, and present these results in Fig. \ref{fig:discussion:sfr_matching}.

As can be seen for the CIGALE runs, as we make the SFR matching criteria stricter, $\mu\Delta$Dust does not appear to change: it remains consistent within error for all CIGALE runs. Indeed, each CIGALE runs remains consistent with $\mu\Delta$Dust~$= 0$. Meanwhile, the uncertainties on $\mu\Delta$Dust increases as a result of the number of merging galaxies that can be studied decreases. This is true for when we do and when we do not include AGN-host galaxies. Therefore, with the limitations of our relatively small sample size, there is no trend of changing $\mu\Delta$Dust with decreasing fraction of $\sigma$ to match within.

For the SIMBA galaxies, the values of $\mu\Delta$Dust decrease as the SFR matching criteria becomes stricter.  Unlike the CIGALE runs, the $\mu\Delta$Dust value is less than zero for matching within $\sigma$/2 when we include AGN-host galaxies, and within all thresholds separations when we exclude AGN hosts.  We interpret the negative $\mu\Delta$Dust as evidence that the destruction and/or removal effects in SIMBA dominate, on average, in the merger sample we studied here. This warrants further, more detailed study that considers other potential influences that are beyond the scope of this work, such as merger stage, gas content, metallicity, and outflows.

\subsection{Dust mass and AGN fraction estimates}
We caution that dust modelling in observations and simulations is notoriously complex. Different spectral energy distribution modelling and fitting tools can have differences in M$_{Dust}$ of $\approx0.3$~dex when using the same dust emission models \citep{2019A&A...621A..51H}. There are many different processes that influence dust creation and destruction, with varying timescales. These processes are also dependent on the properties of the host galaxy beyond those studied in this work. Metallicity of the gas in a galaxy, for example, can influence the rate at which dust is created or destroyed \citep[e.g.][]{2013EP&S...65..213A, 2022MNRAS.509L...6P}. Therefore the metallicity of the galaxies involved in a merger may influence our results but is not studied here due to our use of broadband photometry.

Our dust masses are estimated using optical to FIR data. At the redshifts used in this work, SPIRE data probes the peak of dust emission, hence our requirement that SPIRE or PACS data is present. Where we only have PACS data, we use a 250$\mu$m upper limit of 1.2~mJy to help better constrain the dust emission. As noted in Sect. \ref{sect:data:NEP}, SCUBA-2 850$\mu$m data is available in NEP but its usefulness is limited in this work due to the low angular resolution and difficulty in deblending, meaning cross-matching with our sources will be unreliable. Recent works with ALMA data have also shown that ALMA also cannot fully constrain M$_{Dust}$, especially when metallicity of the galaxies is unknown \citep{2025ApJ...995L..63L}. Therefore, we believe our M$_{Dust}$ estimates are as reliable as we can realistically achieve, as demonstrated in Fig. \ref{fig:data:mdust}, while still not perfect.

In the mid-IR, we are also limited with our data coverage, with only 929 (15.6\%) of our 5941 \textit{Herschel} detected sources having AKARI detections. Only 4.5\%, 4.4\%, 2.1\%, and 2.0\% of matched mergers and non-mergers from the \texttt{cf00\_dr14}, \texttt{cf00\_themis}, \texttt{sb\_themis}, and \texttt{sb\_dr14} runs have AKARI detections when unmatched by SFR (including AGN hosts). This changes to 6.2\%, 5.0\%, 1.9\%, and 1.3\% for the same CIGALE runs, respectively, when matching by SFR. This drop is partially due to a smaller number of AKARI detected sources having reduced $\chi^{2} > 2$, compared to the whole \textit{Herschel} detected sample: depending on the CIGALE run, at most 63\% of AKARI sources have reduced $\chi^{2} > 2$ compared to at least 81\% of all the original 5941 galaxies. Indeed, the fraction of small grains relative to the total M$_{Dust}$ in both star-forming galaxies and post-starbursts is expected to remain below 2-3\% for M$_{\star}$ range probed in this work \citep{2020ApJ...889..150A, 2026arXiv260827571D}. We therefore assume that, while unlikely to have a large impact on M$_{Dust}$, the lack of mid-IR data will mostly affect constraints on \texttt{fracAGN}. So we caution the reader that there is limited constraint on \texttt{fracAGN} in our sample used to calculate $\mu\Delta$Dust.

\section{Conclusions}\label{sect:conclusion}
In this work, we compared the M$_{Dust}$ in merging galaxies to the M$_{Dust}$ in non-merging galaxies. This was done using both observed and simulated galaxies.

We calculated M$_{Dust}$ of over 500 known mergers in NEP using CIGALE. We ran CIGALE using combinations of the \texttt{dustatt\_modified\_CF00} and \texttt{dustatt\_modified\_starburst} dust attenuation models, and the \texttt{dl2014} and \texttt{themis} dust emission modes, giving four runs. We also used the M$_{Dust}$ of nearly 13\,000 galaxy mergers from the SIMBA simulation. We matched the merging systems to five non-merging galaxies with similar M$_{\star}$, local density, and redshift. We also made a further match to include similar SFRs. These matches left approximately 120 observed and 10\,600 SIMBA merging galaxies, when not matching with SFR, or approximately 60 observed and 9600 SIMBA galaxies, when matched with SFR. When we also exclude AGN-host galaxies from our analysis, these matches left approximately 70 observed and 8000 SIMBA merging galaxies, when not matching with SFR, and approximately 35 observed and 7100 SIMBA galaxies, when matched with SFR.

When not matched with SFR, we find that M$_{Dust}$ is increased in observed merging galaxies compared to observed non-merging galaxies by up to a factor of 1.82$\pm$0.35, or 1.60$\pm$0.44 when we exclude AGN-host galaxies. This was for all four CIGALE runs. When matched with SFR, the observed mergers became consistent with no increase in M$_{Dust}$ for all four CIGALE runs. This is true both regardless of including or excluding AGN-host galaxies. Therefore, we find that observed merging galaxies are typically more dusty than observed non-merging galaxies but only because they typically have higher SFRs.

SIMBA galaxies, on the other hand, showed a decrease in M$_{Dust}$ during a merger when we exclude AGN hosts. This decrease did not change when matched to non-merging galaxies with similar SFR. The decrease in M$_{Dust}$ during a merger shows that the SIMBA simulation is missing or mis-modelling some dust processes. It is unclear which processes these are. When we do not exclude AGN hosts, the results from SIMBA are similar to the results from observations: when we match in SFR, the excess in M$_{Dust}$ in mergers decreases but does not become consistent M$_{Dust}$ in non-merging galaxies. This suggests that AGN activity is intertwined with the physical properties of simulated mergers in a way that differs from what we see in observations. This motivates future exploration to disentangle dust, gas, and AGN feedback in simulations.

To test how close the SFR match needs to be before the M$_{Dust}$ enhancement is lost, we used increasingly closely matched SFR to pair the merging galaxies and non-merging systems. As the closeness of SFR became more strict, M$_{Dust}$ did not significantly change for observed galaxies. However, this was dominated by the large uncertainties. The SIMBA galaxies, on the other hand, showed merging galaxies had lower M$_{Dust}$ than non-merging galaxies as the SFR matching became more strict, both when we include and exclude AGN-host galaxies. This again suggests miss-modelling of dust processes in merging galaxies.

As a result of our comparison between observations and simulations, we find that the dust modelling used in SIMBA does not accurately account for the changes in M$_{Dust}$ found in observations. Thus future simulations may need a more refined dust model to accurately capture the observed changes.

This work only uses a simple binary classification: merger or non-merger. By using a more refined non-merger, pre-merger and post-merger classification, we will more closely study when the changes in M$_{Dust}$ occur in the merger sequence. We will also study how these changes in M$_{Dust}$ impact the dust extinction of merging galaxies. Recent Subaru Prime Focal Spectrograph observations in NEP (PI: T.~Goto and PI: W.~J.~Pearson) will also be used to study the impact of merging galaxies on spectroscopically derived dust attenuation and metallicity.

\begin{acknowledgements}
    We would like to thank the referee for their thorough and thoughtful comments that helped improve the quality and clarity of this paper.
    W.J.P., S.D., and A.P.C have been supported by the Polish National Science Center project UMO-2023/51/D/ST9/00147.
    K.M. acknowledges the support of the Polish National Science Center project UMO-2024/53/B/ST9/00230.
    D.D. acknowledges support from the Polish National Agency for Academic Exchange (Bekker grant BPN /BEK/2024/1/00029/DEC/1)
    K.L. acknowledges the support of the National Science Centre, Poland, through the PRELUDIUM grant UMO-2023/49/N/ST9/00746 and support from the GEELSBE2 project with reference PID2023-150393NB-I00 funded by MCIU/AEI/10.13039/501100011033 and the FSE+, and also the Consolidación Investigadora IGADLE project with reference CNS2024-154572. He also gratefully acknowledges financial support of the Department of Education, Junta de Castilla y Le\'on and FEDER Funds (Reference: CLU-2023-1-05).
    S.D. has been supported by the Polish National Science Center project DEC-2025/09/X/ST9/01569.
    T.G. acknowledges the support of  the National Science and Technology Council of Taiwan through grants 114-2112-M-007 -026 -, and 113 -2927-I-007 -501 -.
    This research was conducted under the agreement on joint mobility projects for the years 2026–2027 between the Polish Academy of Sciences and the National Science and Technology Council in Taiwan.
\end{acknowledgements}

\bibliographystyle{aa}
\bibliography{Sonata-DustyMergers}

\begin{appendix}

\section{Deblending of PACS and SPIRE}\label{app:deblend}
To get accurate measurements of the dust content, it is necessary to have reliable far-infrared (FIR) measurements. \textit{Herschel} PACS data, and \textit{Herschel} SPIRE data in particular, are known to suffer from confusion and blending. To deblend these data, we follow \citet{2017A&A...603A.102P}: we predict the flux density in the PACS and SPIRE bands, select sources with predicted flux densities above a certain threshold and use their positions, predicted flux densities, and predicted flux density uncertainties as priors in the Bayesian deblending tool XID+ \citep{2017MNRAS.464..885H, 2017A&A...603A.102P}. We deblended the PACS and SPIRE images provided by the \textit{Herschel} Extragalactic Legacy Project \citep[HELP;][]{2017PKAS...32..219P, 2021MNRAS.507..129S}. We split the PACS and SPIRE images into Hierarchical Equal Area isoLatitude Pixelization of a sphere \citep[HEALPix;][]{2005ApJ...622..759G} tiles with order 10 when running XID+.

We ran CIGALE \citep{2019A&A...622A.103B} to estimate the PACS $100 \mu m$ (PACS green), PACS $160 \mu m$ (PACS red), SPIRE $250 \mu m$ (PSW), SPIRE $350 \mu m$ (PMW), and SPIRE $500 \mu m$ (PLW) fluxes. To estimate these fluxes, we fitted SEDs to CFHT Megacam $u$, HSC $g$, $r$, $i$, $z$, and $Y$, KNPO Flamingos $J$ and $H$, CFHT Wircam $J$, and $Ks$, \textit{Spitzer} IRAC $3.6 \mu m$ and $4.5 \mu m$, and AKARI $2.4 \mu m$ (N2), $3.2 \mu m$ (N3), $4.1 \mu m$ (N4), $7.0 \mu m$ (S7), $9.0 \mu m$ (S9w), $11.0 \mu m$ (S11), $15.0 \mu m$ (L15), $18.0 \mu m$ (L18w), and $24.0 \mu m$ (L24), where available. As discussed in Sect. \ref{sect:data}, the bands were merged based on the positions of the HSC data. The sources that were fitted with CIGALE either had spectroscopic redshifts \citep{2013ApJS..207...37S, 2021MNRAS.502..140H} where available, or photometric redshifts from \citet{2021MNRAS.506.6063H} or \citet{2021MNRAS.502..140H}. We use the \citet{2021MNRAS.506.6063H} redshift where sources have redshifts from both catalogues as \citet{2021MNRAS.506.6063H} provides more photometric redshifts than \citet{2021MNRAS.502..140H}, and the galaxy mergers were selected using \citet{2021MNRAS.506.6063H} redshifts. Otherwise, we used the \citet{2021MNRAS.502..140H} photometric redshifts.

The CIGALE setup is similar to that used by \citet{2017A&A...603A.102P} for the same purpose: an exponentially declining star-formation rate history (SFR) with an exponentially declining burst (\texttt{sfh2exp}), \citet{2003MNRAS.344.1000B} stellar population (\texttt{bc03}) with a \citet{2003PASP..115..763C} initial mass function, a modified \citet{2000ApJ...539..718C} dust attenuation (\texttt{dustatt\_modified\_CF00}), \citet{2014ApJ...780..172D} dust emission (\texttt{dl2014}), and \citet{2006MNRAS.366..767F} active galactic nuclei model (\texttt{fritz2006}) with a \citet{2005AA...437..861S} disk spectrum. The parameters used are presented in Table \ref{tab:app:deblend:cigale}.

\begin{table*}
    \caption{\label{tab:app:deblend:cigale}Parameters used for the CIGALE model SEDs used to estimate PACS and SPIRE fluxes, following \citet{2017A&A...603A.102P} where relevant.}
    \centering
    \begin{tabular}{l c c}
	\hline
	\hline
	Parameter & Value & Description\\
	\hline
	\multicolumn{3}{c}{Star Formation History (\texttt{sfh2exp})} \\
	\hline
	$\tau_{\mathrm{main}}$ & 1000, 3000, 7000 & e-folding time (main)\\
	$\tau_{\mathrm{burst}}$ & 9000, 13000 & e-folding time (burst) \\
	$f_{\mathrm{burst}}$ & 0.001, 0.01, 0.10, 0.20 & Burst mass fraction\\
	Age & 1000, 1329, 1768, 2351, 3126, & Population age (main)\\
            & 4157, 5528, 7352, 9776, 13000 & \\
	Burst Age & 1, 10, 30, 100, 300 & Population age (burst)\\
	& & \\
	\hline
	\multicolumn{3}{c}{Stellar Population \citep{2003MNRAS.344.1000B}}\\
	\hline
	IMF & \citet{2003PASP..115..763C} & Initial Mass Function\\
	$Z$ & 0.02 & Metallicity (0.02 is Solar)\\
	Separation Age & 10 & Separation between young and old stellar populations\\
	& & \\
	\hline
	\multicolumn{3}{c}{Dust Attenuation \citep[Modified ][]{2000ApJ...539..718C}}\\
	\hline
	A$_\mathrm{V}^{\mathrm{BC}}$ & 0.1, 0.18, 0.33, 0.6, 1.1, 2.0 & V-band attenuation of the birth clouds\\
	$\mu$ & 0.5 & Av$_{ISM}$ / (Av$_{BC}$+Av$_{ISM}$)\\
	Slope$_{\mathrm{BC}}$ & -0.7 & Birth cloud attenuation power law slope\\
	Slope$_{\mathrm{ISM}}$ & -1.3 & ISM attenuation power law slope\\
	& & \\
	\hline
	\multicolumn{3}{c}{Dust Emission \citep{2014ApJ...780..172D}}\\
	\hline
	$q_{\mathrm{PAH}}$ & 1.12, 2.50, 3.19 & Mass fraction of PAH \\
	$U_{\mathrm{min}}$ & 5.0, 10.0, 25.0 & Minimum scaling factor of the radiation field intensity\\
	$\alpha$ & 2.0 & Dust power law slope\\
	$\gamma$ & 0.02 & Illuminated fraction\\
	& & \\
	\hline
	\multicolumn{3}{c}{AGN \citep{2006MNRAS.366..767F}}\\
	\hline
	$r_{ratio}$ & 60 & Ratio of the maximum to minimum radii of the dust torus\\
	$\tau$ & 1.0, 6.0 & Optical depth at 9.7 microns\\
	$\beta$ & -0.5 & Beta\\
	$\gamma$ & 0.0 & Gamma\\
	opening angle & 100 & Opening angle of the dust torus\\
	psy & 0.001, 89.990 & Angle between equatorial axis and line of sight\\
	disk type & 1 & Disk spectrum \citep{2005AA...437..861S}\\
	$\delta$ & -0.36 & Power-law of index $\delta$ modifying the optical slop of the disk.\\
	fracAGN & 0.0, 0.05, 0.1, 0.2, 0.3, & AGN fraction\\
         & 0.4, 0.5, 0.6, 0.7 & \\
	$\lambda_{\mathrm{fracAGN}}$ & 0/0 & Wavelength range in microns where to compute the AGN fraction\\
	law & 0 & Extinction law of the polar dust (0 = Small Magellanic Cloud)\\
	EBV & 0.03 & E(B-V) for the extinction in the polar direction in magnitudes\\
	temperature & 100 & Temperature of the polar dust in K\\
	emissivity & 1.6 & Emissivity index of the polar dust\\
	& & \\
	\hline
    
	\end{tabular}
\end{table*}

For the PACS bands, we follow \citet{2024A&A...688A..20W} and select galaxies with predicted PACS 100$\mu m$ flux densities greater than 1~mJy. As not all AKARI sources matched to the optical HSC positions, suggesting these sources are dust obscured but are likely to be visible in the FIR, we also include the sources with detections in at least one of the S7, S9w, S11, L15, L18w, or L24 bands. As these AKARI sources do not have flux density predictions for the PACS bands, we follow \citet{2017A&A...603A.102P} and give the AKARI sources a predicted flux density of 7~mJy and flux density uncertainty of 1e7~mJy to produce a functionally flat prior. Due to the relatively small PACS coverage, compared to the size of the NEP field, we limit the prior catalogue to sources that fall within the region bounded by the RA-Dec positions (269.78\degree, 66.86\degree), (267.94\degree, 66.29\degree), (267.94\degree, 66.39\degree), and (268.25\degree, 66.99\degree); the area that PACS covered. As XID+ requires images with units of Jy/beam and the PACS images have units of MJy/sr, we  multiply the PACS 100$\mu m$ and PACS 160$\mu m$ images by 0.017 and 0.048, respectively. This results in 27\,180 sources with PACS 100$\mu m$ and PACS 160$\mu m$ flux densities. We multiply the extracted fluxes by a factor of 1.68 and 1.49 in the 100$\mu m$ and 160$\mu m$ bands, respectively, to correct for the aperture \citep{2019PASJ...71...13P}.

For the SPIRE bands, we deviate from \citet{2017A&A...603A.102P} and select galaxies with predicted PSW flux densities greater than 1.25~mJy and again include missing AKARI sources. The SPIRE flux densities found by exactly following \citet{2017A&A...603A.102P} are known to slightly suffer from over-deblending \citep{2019A&A...631A..51P}, likely caused by having too many sources in the prior list for XID+ meaning the flux is spread between sources in a non-physical way. \citet{2017A&A...603A.102P} had approximately and average of 3.5 sources per PSW beam in their prior catalogue with a flux density cut of 0.7~mJy. Using our cut of 1.25~mJy, this produces an average of approximately 2 sources per PSW beam, including the missing AKARI sources. This results in 300\,263 sources with PSW, PMW, and PLW flux densities. The SPIRE observations cover a larger region than the PACS observations, meaning that despite a higher source density for PACS, there are more sources with SPIRE flux densities.

Both the PACS and SPIRE images contain the Cat's Eye nebula. XID+ is unable to reliably process extended sources, meaning regions close to extended sources will not have reliable flux densities. As a result, we remove the sources that are within a HealPix tile of order 10 that also contains the Cat's Eye nebula (HEALPix tiles 2\,970\,995, 2\,970\,998, 2\,971\,001, and 2\,971\,004). This removes 345 PACS sources and 656 SPIRE sources, leaving 26\,835 sources with PACS flux densities and 299\,607 sources with SPIRE flux densities.

XID+ reports the 32nd and 68th percentile flux densities, with the differences between these and the 50th percentile flux density giving the lower and upper 1$\sigma$ uncertainties. The resulting errors are known to be underestimated \citep{2024A&A...688A..20W} and so we use the difference between the 32nd and 68th percentiles as the error, effectively doubling the estimated uncertainty \citep{2024A&A...688A..20W}. For the NEP-HSC galaxies that were not selected to be in the PACS prior catalogue for XID+ but lay within the PACS region described above, we set the upper limit of PACS green to be 1~mJy for further use. The NEP-HSC galaxies not in the SPIRE prior catalogue has their upper 250$\mu m$ band limit set to 1.25~mJy. The PACS red, and SPIRE 350$\mu m$ and 500$\mu m$ bands did not have an upper limit set.

\FloatBarrier
\section{CIGALE model parameters}\label{app:models}
\FloatBarrier
Here we present the parameters used for the SFH and stellar population models used in the four CIGALE runs in Table \ref{table:app:models:sfh}. We also present the parameters used for the two dust attenuation and two dust emission models used in the CIGALE runs in Table \ref{table:app:models:dust}.

\begin{figure*}
    \centering
    \resizebox{0.91\hsize}{!}{\includegraphics{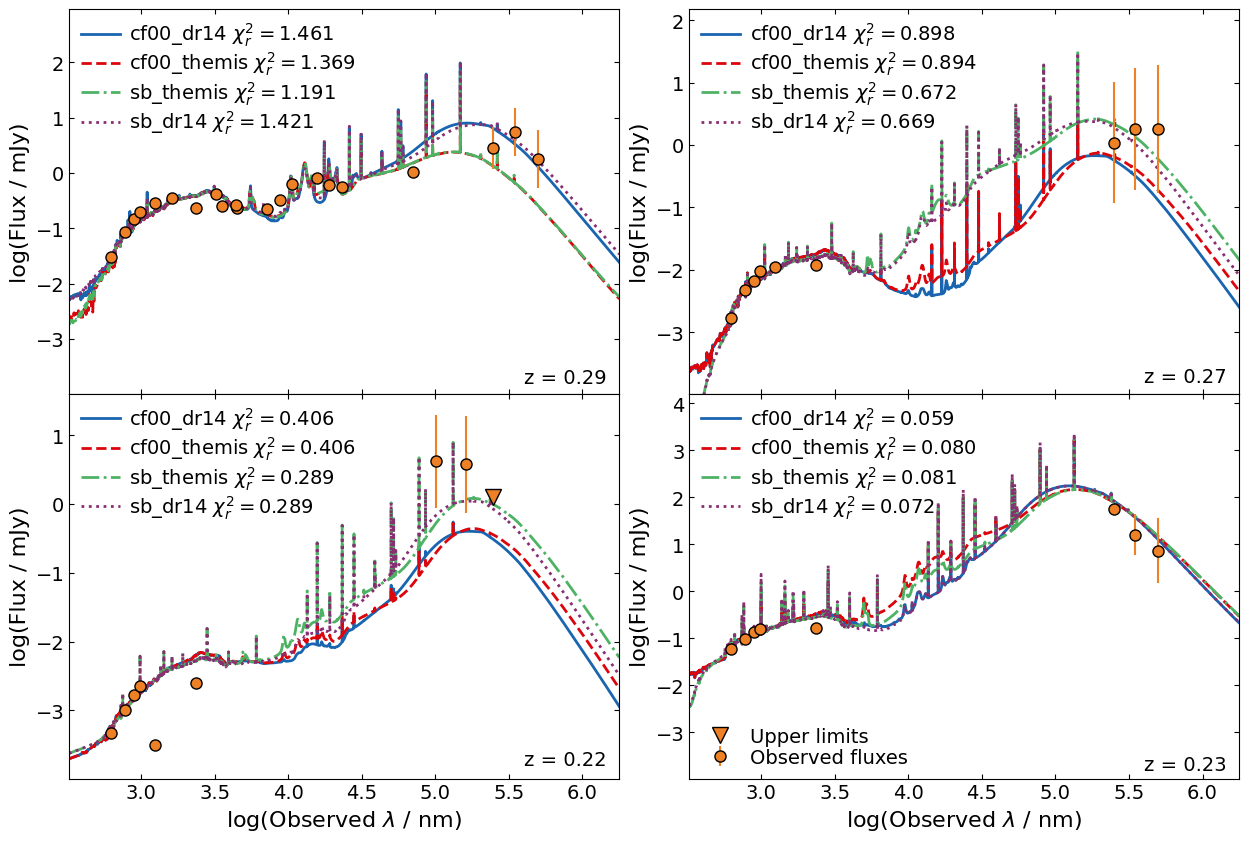}}
    \caption{Best fitting SEDs from the \texttt{cf00\_dr14} (solid blue line), \texttt{cf00\_themis} (dashed red line), \texttt{sb\_themis} (dot-dashed green line), and \texttt{sb\_dr14} (dotted purple line) CIGALE runs to the photometric detections (orange circles) or upper limits (orange triangles) for four observed galaxies. The reduced $\chi^{2}$ ($\chi^{2}_{r}$) for each of the CIGALE runs for each galaxy is shown, as well as the redshift of the galaxy.}
    \label{fig:app:models:sed}
\end{figure*}

We also present the best fitting SEDs for each of the four CIGALE runs for four galaxies in Fig. \ref{fig:app:models:sed}. These galaxies were randomly selected from all the galaxies that have reduced $\chi^{2} < 2$ in all four CIGALE runs.

\begin{table*}
    \caption{\label{table:app:models:sfh}Parameters for SFH, stellar population, and AGN models used for the four CIGALE runs to estimate dust mass.}
    \centering
    \begin{tabular}{l c c}
	\hline
	\hline
	Parameter & Value & Description\\
	\hline
	\multicolumn{3}{c}{Star Formation History (\texttt{sfh2exp})} \\
	\hline
	$\tau_{\mathrm{main}}$ & 1000, 3000, 7000 & e-folding time (main)\\
	$\tau_{\mathrm{burst}}$ & 9000, 13000 & e-folding time (burst) \\
	$f_{\mathrm{burst}}$ & 0.001, 0.01, 0.10, 0.20 & Burst mass fraction\\
	Age & 1000, 1329, 1768, 2351, 3126, & Population age (main)\\
            & 4157, 5528, 7352, 9776, 13000 & \\
	Burst Age & 1, 10, 30, 100, 300 & Population age (burst)\\
	\hline
	\multicolumn{3}{c}{Stellar Population \citep{2003MNRAS.344.1000B}}\\
	\hline
	IMF & \citet{2003PASP..115..763C} & Initial Mass Function\\
	$Z$ & 0.02 & Metallicity (0.02 is Solar)\\
	Separation Age & 10 & Separation between young and old stellar populations\\
	\hline
        \multicolumn{3}{c}{Nebular emission}\\
        \hline
        logU & -2.0 & Ionisation parameter\\
        zgas & 0.02 & Gas metallicity\\
        ne & 100 & Electron density\\
        f$_{esc}$ & 0.0 & Fraction of Lyman continuum photons escaping the galaxy\\
        f$_{dust}$ & 0.0 & Fraction of Lyman continuum photons absorbed by dust\\
        lines\_width & 300.0 & Line width in kms$^{-1}$\\
        emission & True & Include nebular emission\\
        \hline
        \multicolumn{3}{c}{AGN model \citep{2016MNRAS.458.2288S})}\\
        \hline
        t & 7 & Average edge-on optical depth at 9.7$\mu$m\\
        pl & 1.0 & Power-law exponent that sets radial gradient of dust density\\
        q & 1.0 & Index that sets dust density gradient with polar angle\\
        oa & 40 & Angle measured between the equatorial plane and edge of the torus\\
        R & 20 & Ratio of outer to inner radius\\
        Mcl & 0.97 & Fraction of total dust mass inside clumps\\
        i & 30 & inclination angle\\
        disk type & 1 & Disk spectrum \citep{2005AA...437..861S}\\
        fracAGN & 0.0, 0.1, 0.2, 0.3, 0.4, 0.5 & AGN fraction\\
        $\lambda_{\mathrm{fracAGN}}$ & 0/0 & Wavelength range in microns where to compute the AGN fraction\\
	law & 0 & Extinction law of the polar dust (0 = Small Magellanic Cloud)\\
	EBV & 0.03 & E(B-V) for the extinction in the polar direction in magnitudes\\
	temperature & 100 & Temperature of the polar dust in K\\
	emissivity & 1.6 & Emissivity index of the polar dust\\
        \hline
        
	\end{tabular}
\end{table*}

\begin{table*}
    \caption{\label{table:app:models:dust}Parameters for dust attenuation and emission models used for the four CIGALE runs to estimate dust mass.}
    \centering
    \begin{tabular}{l c c}
	\hline
	\hline
	Parameter & Value & Description\\
	\hline
    
	\multicolumn{3}{c}{Dust Attenuation \citep[Modified ][]{2000ApJ...539..718C}}\\
	\hline
	A$_\mathrm{V}^{\mathrm{BC}}$ & 0.1, 0.18, 0.33, 0.6, 1.1, 2.0 & V-band attenuation of the birth clouds\\
	$\mu$ & 0.5 & Av$_{ISM}$ / (Av$_{BC}$+Av$_{ISM}$)\\
	Slope$_{\mathrm{BC}}$ & -0.7 & Birth cloud attenuation power law slope\\
	Slope$_{\mathrm{ISM}}$ & -1.3 & ISM attenuation power law slope\\

        \hline
        \multicolumn{3}{c}{Dust Attenuation \citep[Modified ][]{2000ApJ...533..682C}}\\
	\hline
	E$_{BV, lines}$ & 0.1, 0.3, 0.5, 0.7, 0.9 & Colour excess of the nebular lines light\\
        E$_{BV, factor}$ & 0.3, 0.5, 0.8, 1.0 & Reduction factor to apply on E$_{BV, lines}$ to compute E(B-V)\\
        UV bump wavelength & 217.5 & Central wavelength of the UV bump in nm\\
        UV bump width & 35.0 & FWHM of the UV bump in nm\\
        UV bump amplitude & 3.0 & Amplitude of the UV bump (3 = Milky Way)\\
        \multirow{2}{*}{Powerlaw slope} & \multirow{2}{*}{-1.3, -0.7, 0.0} & Slope delta of the power law modifying the\\
         & & attenuation curve\\
        \multirow{2}{*}{Ext law$_{emission lines}$} & \multirow{2}{*}{1} & Extinction law to use for attenuating the emission\\
         & & lines flux (1 = Milky Way)\\
	\hline
        \hline
    
	\multicolumn{3}{c}{Dust Emission \citep{2014ApJ...780..172D}}\\
	\hline
	$q_{\mathrm{PAH}}$ & 0.47, 1.12, 2.50, 3.9, 5.26, 6.63 & Mass fraction of PAH \\
	$U_{\mathrm{min}}$ & 5.0, 10.0, 25.0 & Minimum scaling factor of the radiation field intensity\\
	$\alpha$ & 2.0 & Dust power law slope\\
	$\gamma$ & 0.01, 0.021, 0.045, 0.095, 0.2 & Illuminated fraction\\
	\hline
    
        \multicolumn{3}{c}{Dust Emission \citep{2017A&A...602A..46J}}\\
	\hline
	  $q_{\mathrm{HAC}}$ & 0.02, 0.10, 0.17, 0.28, 0.40 & Mass fraction of hydrocarbon solids \\
	$U_{\mathrm{min}}$ & 5.0, 10.0, 25.0 & Minimum radiation field\\
	$\alpha$ & 2.0 & Dust power law slope\\
	$\gamma$ & 0.01, 0.021, 0.045, 0.095, 0.2 & Illuminated fraction\\
	\hline
    
	\end{tabular}
\end{table*}

\twocolumn

\FloatBarrier
\section{$\mu\Delta$Dust including non-FIR detected sources}\label{app:notFIRonly}
\begin{figure}
    \centering
    \resizebox{\hsize}{!}{\includegraphics{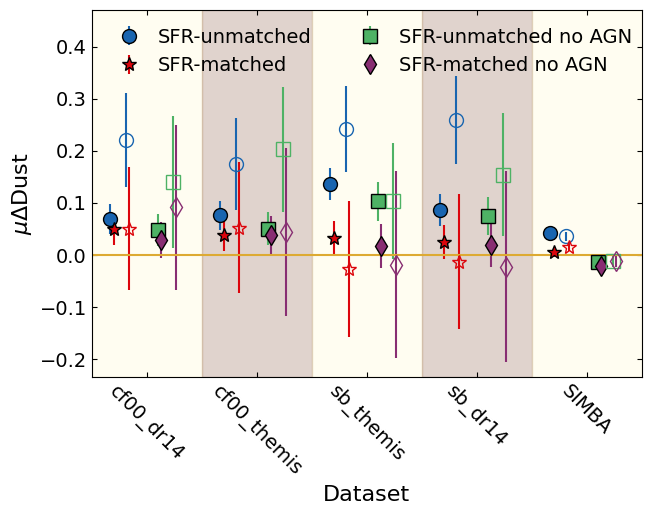}}
	\caption{As Fig. \ref{fig:results:main} but including galaxies without FIR detections. Unfilled markers indicate the results from Fig. \ref{fig:results:main} for reference.}
	\label{fig:app:notFIRonly:results}
\end{figure}

\FloatBarrier
Due to our limited FIR coverage potentially influencing our results, we repeat our analysis using all sources, including sources that are not detected in the \textit{Herschel} bands (PACS and/or SPIRE). For sources without SPIRE detections, we apply a 250$\mu$m upper limit of 1.2~mJy during SED fitting. We also repeat our analysis for all SIMBA galaxies. We present the number of merging galaxies that could be matched to five non-merging galaxies for each of the four CIGALE runs in Table \ref{table:app:notFIRonly:numbers}. The $\mu\Delta$Dust values found using these galaxies are presented in Table \ref{table:app:notFIRonly:results}. We also include these values for when we also exclude AGN host galaxies in the same tables. The $\mu\Delta$Dust values are also presented in Fig. \ref{fig:app:notFIRonly:results}.

The $\mu\Delta$Dust values when we do not match the mergers to non-mergers in SFR are much lower than the values found when we only allow galaxies with \textit{Herschel} detections. When we further match the non-mergers by SFR, this dust mass enhancement is not as strong for the runs with \texttt{dustatt\_modified\_CF00}, or consistent with no increase in dust mass for the runs with \texttt{dustatt\_modified\_starburst}. However, due to the much larger sample size, the uncertainties in these values are much smaller than when we only allow FIR detected sources. For the SIMBA galaxies when not matching in SFR, the $\mu\Delta$Dust value is larger than, but consistent with, the $\mu\Delta$Dust value when only using FIR detectable sources. $\mu\Delta$Dust becomes consistent with zero when we match the SIMBA mergers in SFR to their control galaxies.

When we exclude AGN host galaxies, the $\mu\Delta$Dust values are consistent with the values we find when we require \textit{Herschel} detections, except for the \texttt{cf00\_themis} run. This agreement is likely a result of the large uncertainties caused by the relatively small sample size when we enforce FIR detections. Much like the results when we include AGN hosts, when we also match the non-mergers in SFR, the $\mu\Delta$Dust values are consistent with zero. For the SIMBA galaxies, we again find that mergers have less dust than non-mergers. However, unlike the FIR detectable sources, this decrement of dust in mergers becomes larger when we match in SFR.

The results when only using FIR detected sources is qualitatively in agreement with the results we find when also allowing galaxies with only 250$\mu$m upper-limits: when not matching by SFR we see larger M$_{Dust}$ in merging galaxies than non-merging galaxies and when we match by SFR, this increase is reduced, or lost. This is for both when we include and exclude AGN host galaxies. However, the much smaller uncertainties we find when including galaxies without FIR detections are likely primarily driven by the smaller number of galaxies. We also caution that the M$_{Dust}$ we derive from galaxies without \textit{Herschel} detections are unlikely to be as reliable as the M$_{Dust}$ for the galaxies presented in the main body of this paper. This is due to the limited constraint on the peak of the dust emission for the majority of the galaxies presented in this Appendix. Therefore, care must be taken to not over-analyse the results presented in this Appendix.

\begin{table}
	\caption{Number of merging galaxies used from each CIGALE run when unmatched and matched in SFR and when including and not including AGN-host galaxies. Not all galaxies have FIR detections.}
	\label{table:app:notFIRonly:numbers}
	\centering
	\begin{tabular}{cccc}
		\hline
        \hline
        \multirow{2}{*}{Dataset} & \multicolumn{2}{c}{Number of Mergers} & \multirow{2}{*}{AGN}\\
         & Unmatched SFR & Matched SFR & \\
        \hline
        \texttt{cf00\_dr14} & 1078 & 906 & yes \\
        \texttt{cf00\_themis} & 1067 & 872 & yes \\
        \texttt{sb\_themis} & 765 & 580 & yes \\
        \texttt{sb\_dr14} & 783 & 620 & yes \\
        SIMBA & 11\,858 & 10\,701 & yes \\
        \hline
        \texttt{cf00\_dr14} & 906 & 767 & no \\
        \texttt{cf00\_themis} & 892 & 734 & no \\
        \texttt{sb\_themis} & 517 & 388 & no \\
        \texttt{sb\_dr14} & 536 & 423 & no \\
        SIMBA & 9207 & 8221 & no\\
        \hline
	\end{tabular}
\end{table}

\begin{table}
	\caption{$\mu\Delta$Dust values for the four CIGALE runs when unmatched and matched in SFR and when including and not including AGN-host galaxies. Not all galaxies have FIR detections.}
	\label{table:app:notFIRonly:results}
	\centering
	\begin{tabular}{cccc}
		\hline
        \hline
        \multirow{2}{*}{Dataset} & \multicolumn{2}{c}{$\mu\Delta$Dust} & with\\
         & Unmatched SFR & Matched SFR & AGN \\
        \hline
        \texttt{cf00\_dr14} & 0.069$\pm$0.029 & 0.049$\pm$0.031 & yes \\
        \texttt{cf00\_themis} & 0.076$\pm$0.028 & 0.039$\pm$0.032 & yes \\
        \texttt{sb\_themis} & 0.137$\pm$0.03 & 0.032$\pm$0.033 & yes \\
        \texttt{sb\_dr14} & 0.086$\pm$0.031 & 0.024$\pm$0.032 & yes \\
        SIMBA & 0.043$\pm$0.009 & 0.005$\pm$0.008 & yes \\
        \hline
        \texttt{cf00\_dr14} & 0.048$\pm$0.031 & 0.029$\pm$0.034 & no \\
        \texttt{cf00\_themis} & 0.051$\pm$0.032 & 0.038$\pm$0.037 & no \\
        \texttt{sb\_themis} & 0.103$\pm$0.038 & 0.017$\pm$0.042 & no \\
        \texttt{sb\_dr14} & 0.075$\pm$0.037 & 0.018$\pm$0.042 & no \\
        SIMBA & -0.013$\pm$0.011 & -0.022$\pm$0.01 & no \\
        \hline
	\end{tabular}
\end{table}

\end{appendix}
\end{document}